\documentclass[longauth]{aa} 
\usepackage{epsfig}     
\usepackage{threeparttable}  
\usepackage{booktabs}
\usepackage{lscape}
\usepackage{multirow}
\usepackage[utf8]{inputenc}
\usepackage{graphicx}
\usepackage{xcolor}
\usepackage{txfonts}
\usepackage{gensymb}
\usepackage{soul} 
\newcommand{\dd}{{\rm d}}
\usepackage{textcomp} 
\begin{document}

\title{First year of energetic particle measurements in the inner heliosphere with Solar Orbiter's Energetic Particle Detector}

\titlerunning{First year of EPD observations}
\authorrunning{Wimmer-Schweingruber et al.}

\author{
R.~F.~Wimmer-Schweingruber\inst{\ref{cau}} \and 
N.~P.~Janitzek\inst{\ref{esac}} \and 
D.~Pacheco\inst{\ref{cau},\ref{corr}} \and  
I.~Cernuda\inst{\ref{uah}} \and 
F.~Espinosa Lara\inst{\ref{uah}} \and 
R.~G\'omez-Herrero\inst{\ref{uah}} \and
G.~M.~Mason\inst{\ref{apl}} \and
R.~C.~Allen\inst{\ref{apl}} \and
Z.~G.~Xu\inst{\ref{cau}} \and
F.~Carcaboso\inst{\ref{uah}} \and
A.~Kollhoff\inst{\ref{cau}} \and
P.~Kühl\inst{\ref{cau}} \and
J.~L.~Freiherr von Forstner\inst{\ref{cau}, \ref{16}} \and
L.~Berger\inst{\ref{cau}} \and
J.~Rodriguez-Pacheco\inst{\ref{uah}} \and
G.~C.~Ho\inst{\ref{apl}} \and
G.~B.~Andrews\inst{\ref{apl}} \and
V.~Angelini\inst{\ref{11}} \and 
A.~Aran\inst{\ref{15}} \and 
S.~Boden\inst{\ref{cau}, \ref{5}} \and
S.~I.~B\"ottcher\inst{\ref{cau}} \and
A.~Carrasco\inst{\ref{uah}} \and
N.~Dresing\inst{\ref{utu}} \and 
S.~Eldrum\inst{\ref{cau}} \and
R.~Elftmann\inst{\ref{cau}} \and
V. Evans\inst{\ref{11}} \and 
O.~Gevin\inst{\ref{9}} \and 
J.~Hayes\inst{\ref{apl}} \and
B.~Heber\inst{\ref{cau}} \and
T.~S.~Horbury\inst{\ref{11}} \and 
S.~R.~Kulkarni\inst{\ref{cau}, \ref{6}} \and
D.~Lario\inst{\ref{14}} \and 
W.~J.~Lees\inst{\ref{apl}} \and
O.~Limousin\inst{\ref{9}} \and 
O.~Malandraki\inst{\ref{12}} \and 
C.~Mart\'in\inst{\ref{cau}, \ref{7}} \and
H.~O’Brien\inst{\ref{11}} \and 
M.~Prieto Mateo\inst{\ref{uah}} \and
A.~Ravanbakhsh\inst{\ref{cau}, \ref{mps}} \and
O.~Rodriguez-Polo\inst{\ref{uah}} \and
S.~S\'anchez Prieto\inst{\ref{uah}} \and
C.~E.~Schlemm\inst{\ref{apl}} \and
H.~Seifert\inst{\ref{apl}} \and
J.~C.~Terasa\inst{\ref{cau}} \and
K.~Tyagi\inst{\ref{apl}, \ref{10}} \and
R.~Vainio\inst{\ref{utu}} \and
A.~Walsh\inst{\ref{esac}} \and
M.~K.~Yedla\inst{\ref{cau}, \ref{mps}}
}

\institute{Institute of Experimental and Applied Physics, Kiel University, 24118 Kiel, Germany\label{cau}
\and
European Space Astronomy Center, Villanueva de la Ca\~nada, 28692 Madrid, Spain\label{esac} \and
Corresponding author, e-mail:pacheco@physik.uni-kiel.de\label{corr} \and
Universidad de Alcal\'a, Space Research Group, 28805 Alcal\'a de Henares, Spain\label{uah}\and 
Johns Hopkins University Applied Physics Laboratory, Laurel, MD, USA \label{apl}\and 
Now at Paradox Cat GmbH, Brienner Str. 53, 80333 M\"{u]}nchen , Germany\label{16}\and
Department of Physics, Imperial College London, London SW7 2AZ, UK \label{11}\and
Departament de Física Qu\`antica i Astrof\'isica, Institut de Ci\`encies del Cosmos (ICCUB), Universitat de Barcelona (UB-IEEC), Barcelona, Spain \label{15}\and
Now at DSI Datensicherheit GmbH, Rodendamm 34, 28816 Stuhr\label{5} \and
Department of Physics and Astronomy, University of Turku, Finland\label{utu}\and
D\'epartement d'Astrophysique, Commissariat à l'\'energie atomique et aux \'energies alternatives, CEA, Saclay, France\label{9}\and
Now at Deutsches Elektronen-Synchrotron (DESY), Platanenallee 6, 15738 Zeuthen, Germany\label{6} \and
Heliophysics Science Division, NASA Goddard Space Flight Center, Greenbelt, MD 20771, USA\label{14}\and
National Observatory of Athens, Institute for Astronomy, Astrophysics, Space Applications and Remote Sensing, Athens, Greece \label{12}\and
Now at German Aerospace Center (DLR), Berlin, Germany. Department of Extrasolar Planets and Atmospheres\label{7} \and
Now at Max-Planck-Institute for Solar System Research, Justus-von-Liebig-Weg 3, 37077 Göttingen, Germany\label{mps}\and
now at Univ. Colorado/LASP, Boulder, CO, USA\label{10}
}

\date{Received ; accepted , }
  \abstract
   {Solar Orbiter strives to unveil how the Sun controls and shapes the heliosphere and fills it with energetic particle radiation. To this end, its Energetic Particle Detector (EPD) has now been in operation, providing excellent data, for just over a year.}
   {EPD measures suprathermal and energetic particles in the energy range from a few keV up to (near-) relativistic energies (few MeV for electrons and about 500 MeV/nuc for ions). We present an overview of the initial results from the first year of operations and we provide a first assessment of issues and limitations. In addition, we present areas where EPD excels and provides opportunities for significant scientific progress in understanding how our Sun shapes the heliosphere.}
   {We used the solar particle events observed by Solar Orbiter on 21 July  and between 10-11 December 2020 to discuss the capabilities, along with updates and open issues related to EPD on Solar Orbiter. We also give some words of caution and caveats related to the use of EPD-derived data.}
   {During this first year of operations of the Solar Orbiter mission, EPD has recorded several particle events at distances between 0.5 and 1 au from the Sun. We present dynamic and time-averaged energy spectra for ions that were measured with a combination of all four EPD sensors, namely:\ the SupraThermal Electron and Proton sensor (STEP), the Electron Proton Telescope (EPT), the Suprathermal Ion Spectrograph (SIS), and the High-Energy Telescope (HET) as well as the associated energy spectra for electrons measured with STEP and EPT. We illustrate the capabilities of the EPD suite using the 10-11 December 2020 solar particle event. This event showed an enrichment of heavy ions as well as $^3$He, for which we also present dynamic spectra measured with SIS. The high anisotropy of electrons at the onset of the event and its temporal evolution is also shown using data from these sensors. We discuss the ongoing in-flight calibration and a few open instrumental issues using data from the 21 July and the 10-11 December 2020 events and give guidelines and examples for the usage of the EPD data. We explain how spacecraft operations may affect EPD data and we present a list of such time periods in the appendix. A list of the most significant particle enhancements as observed by EPT during this first year is also provided.} 
   {}
   \maketitle
%
\section{Introduction}\label{sec_intro}

Understanding how the Sun accelerates particles to relativistic energies and how these propagate from their acceleration site to fill the heliosphere are among the key questions that Solar Orbiter has set out to answer \cite[]{Mueller_2020}. As one of the ten instruments on board Solar Orbiter, the Energetic Particle Detector (EPD) is key in measuring energetic particles and thereby solving this riddle \citep{RodriguezPacheco_2020}. EPD consists of four individual sensors that are served by a common Instrument Control Unit (ICU). The Supra-Thermal Electron Proton sensor (STEP) measures electrons and ions with energies between 4 keV ($\sim 6$ keV for ions) and 80 keV with high pitch-angle resolution around the direction of nominal magnetic connection to the Sun. The Electron Proton Telescope (EPT) shares its electronics with the High-Energy Telescope (HET) and measures electrons and ions between 25 keV and 400 keV (6.9 MeV and beyond for ions, as discussed below) in four observational directions, which are also shared with HET. Neither STEP nor EPT discriminate between different ion species.

HET measures electrons from 300 keV to 30 MeV and ions from 6.8 MeV to more than 100 MeV/nuc, with the upper energy limit depending on the ion species.
HET can discriminate between different elements and the upper energy range depends on the nuclear charge. Finally, the Suprathermal Ion Spectrograph (SIS) measures the elemental and isotopic composition of ions in the energy range from approximately 14 keV/nuc to $\sim$ 20.5 MeV/nuc with its two large-aperture viewing directions. Thus, EPD covers four viewing directions with EPT/HET, one with STEP, and two with SIS. The EPD instrument is described in more detail in \cite{RodriguezPacheco_2020}, where the energy ranges for each sensor and particle species are included, along with the sensors' field of view; henceforth, we refer to that paper as the ``instrument paper.'' The information presented in this article is meant to complement the instrument paper with the most up-to-date information. Additional engineering and testing information is given in \citet{prieto-etal-2021}. Obviously, pitch-angle information can only be derived with the invaluable data from the Solar Orbiter magnetometer \cite[MAG,][]{Horbury_2020}. 

In this first-year overview paper, we provide updates on EPD as well as additional insights into its operation and data products, which are based on EPD's first year of operation. The aim is to give the reader a good understanding of the data and data products provided by EPD and to point out possible pitfalls and open issues. The format chosen for this is to present a selection of observations from the time period between 28 February 2020 and 28 February 2021 to illustrate EPD data products and provide pertinent information that was not yet available when the instrument paper was written.
Solar activity was low in this first year of operations and solar particle energies only rarely exceeded the lower energy threshold of HET. This, in turn, means that we have not yet had the chance to discover the full extent of the EPD instrument. For instance, as we are only beginning to exit the deep solar minimum between solar cycles 24 and 25, so far we have not encountered an event with the very high counting rates that are typically associated with the largest solar energetic particle (SEP) events. 

This paper is divided into seven sections. The following section (Sect.~\ref{sec:overview}) provides an overview of the measurements discussed in this paper, namely from 28 February 2020 to 28 February 2021.
In Sect.~\ref{sec:Dec_event}, we discuss in more detail a particle event observed in December 2020 to illustrate the capabilities of EPD, but also some of the potential pitfalls lurking in the depths of EPD's data products. While EPD was calibrated before flight \citep{RodriguezPacheco_2020} certain properties have to be verified or refined in an in-flight calibration in space. This in-flight calibration, performed during the first year of EPD operation, is discussed in Sect.~\ref{sec:in_flight_calib}.
We also present aspects of EPD which were not yet known at the time of the instrument paper's publication, and we describe how we have addressed them in the published data. Using the 21 July 2020 event, Sect.~\ref{subsec:EPDsensor_comparison} describes how the full EPD energy spectra of heavy-ion-rich events can be analyzed in a consistent way by combining data from EPT, SIS, and HET. 
There are a number of spacecraft operations that can affect EPD data; these are presented in Sect.~\ref{sec:ops}. A list of all time periods affected by spacecraft operations is given in Appendix~\ref{app:ops}. Finally, our conclusions and an outlook are given in Section~\ref{sec:conclusions}. Appendix~\ref{app:events} contains a list of all particle enhancements observed in the first year of EPD operations.

\section{Overview of EPD measurements} \label{sec:overview}

\begin{figure*}[h!t]
    \centering
    \includegraphics[width=\textwidth]{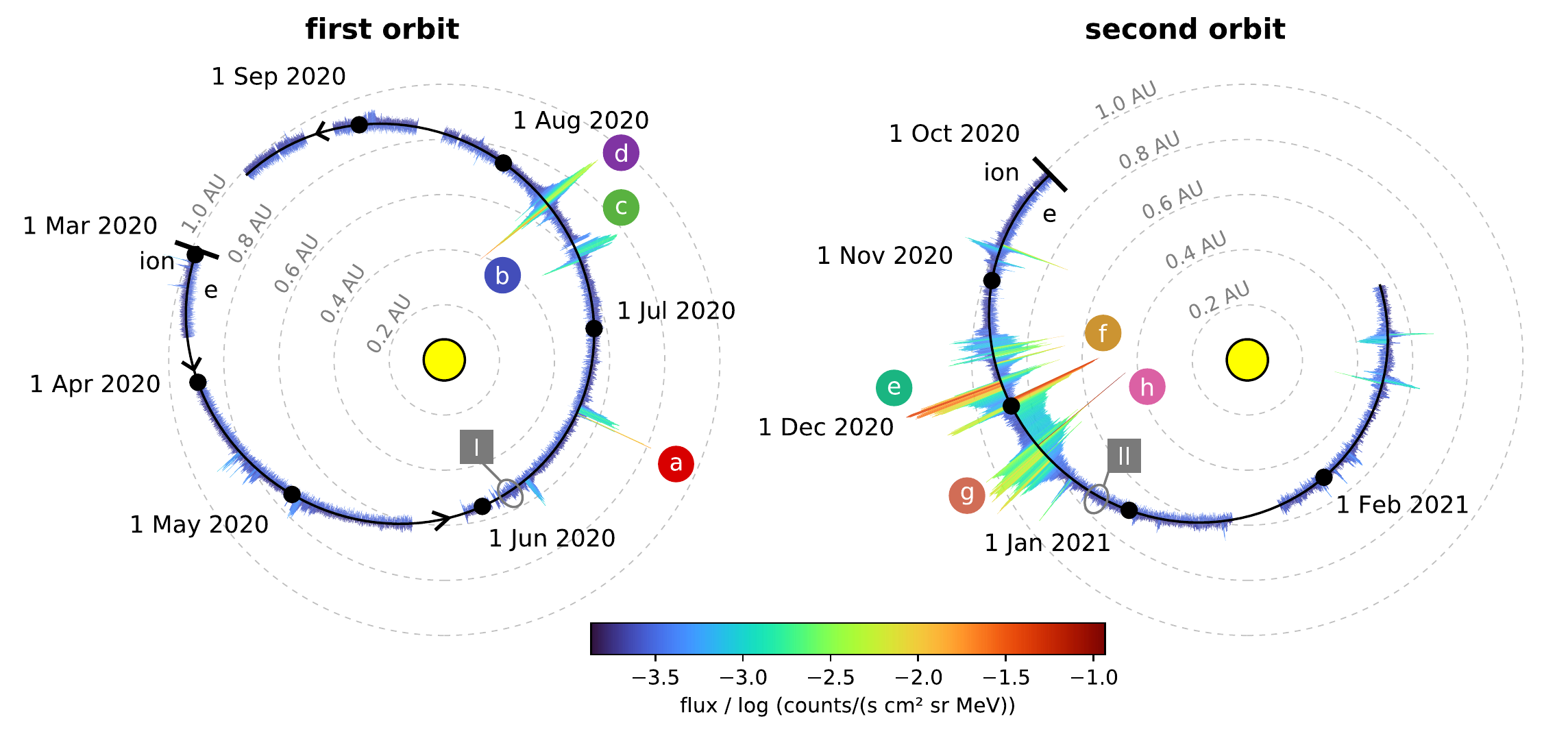}
    \caption{One year's worth of data from EPD/EPT. Intensities, in the energy range of 124--218 keV for ions (outer) and 54--101 keV for electrons (inner), are color-coded along Solar Orbiter's orbit, shown in the mean ecliptic and equinox of J2000 (ECLIPJ2000) reference frame. The left-hand panel shows data for the first orbit (28 February to 1 October 2020), the right hand panel shows data from the second (currently incomplete) orbit (1 October 2020 to 28 February 2021, right after the second mission perihelion). Data gaps are due to commissioning activities. Interesting periods are labeled: with "a" through "h" referencing event studies; and "I" and "II" indicating the encounter with the tail of the comet Atlas and the Venus fly-by, respectively.}
    \label{fig:orbits}
\end{figure*}

Commissioning of the STEP, EPT, and HET sensors began on 28 February 2020, when EPD was switched on (with the exception of SIS which required more time to outgas before its high voltages were turned on). The formal commissioning phase ended in mid-June 2020, however, EPD has been providing valuable data well before that date. For example, \cite{vforstner-etal-2021} used EPD data from April 2020 to investigate the Forbush decrease caused by the 19-20 April interplanetary coronal mass ejection (ICME) and analyze its propagation from 0.2 au upstream of Earth to the Earth Lagrange point L1, and \citet{kilpua-etal-2021} include EPD data in their analysis of the same event. \citet{telloni-etal-2021} investigate the interaction of two ICMEs on 7--8 June 2020, thus illustrating the scientific value of data even during the commissioning phase. Because EPD data from its commissioning phase is scientifically useful, this paper reports on the first year's worth of data from EPD and Solar Orbiter from 28 February 2020 to 28 February 2021. The beginning of Solar Orbiter's cruise phase was marked by the 18-20 June ion event seen by both STEP and EPT, as well as by SIS and the C detectors of HET, which is investigated by \citet{aran-etal-2021}.

Figure~\ref{fig:orbits} shows Solar Orbiter's trajectory for the first year of measurements of EPD, from 28 February 2020 to 28 February 2021 with intensities of 54--101 keV electrons on the inner side and 124\,--\,218 keV ions on the outer side as measured by EPT. The figure shows the projection of the orbit onto the ecliptic of J2000 (ECLIPJ2000). The two orbits shown cover the time period reported here. Apart from a low level of background activity \citep[see][for an in-depth discussion]{mason-etal-2021} and small increases due to corotating interaction regions \citep{allen-etal-2020}, a number of intensity increases or particle events can be seen - quantitative information is color-coded, but also given by the extent away from the orbital line. Their main characteristics, such as onset and peaking times, are listed in Tables~\ref{tab:catalogue_electrons} (for electrons) and \ref{tab:catalogue_ions} (for ions).

Several particle events shown in Figure~\ref{fig:orbits} are discussed in more detail in accompanying papers in this special issue: the small ion event seen between 18-20 June 2020, indicated by (a) in Fig.~\ref{fig:orbits}, is discussed by \citet{aran-etal-2021}, who find that the lack of electrons and type III radio bursts and the simultaneous response of the ion intensity-time profiles at various energies indicates an interplanetary source for the particles.
Over July 2020, a series of intense near-relativistic electron events were measured by EPD, indicated by (b) in Fig.~\ref{fig:orbits}, accompanied by ion components in some cases. \citet{gomez-herrero-etal-2021} investigated these events, their solar origin and the conditions for the interplanetary transport of the particles. 
\citet{Mason_2020} analyzed a series of five impulsive ion events between 18 June and 17 September, shown in (c), which covers the time of the first perihelion pass of Solar Orbiter. These events could be identified as $^3$He-rich impulsive events with associated type III bursts. They provide a detailed study of heavy ion spectra and composition for the event on 20-21 July, indicated by (d) in Fig.~\ref{fig:orbits}, which illustrates the excellent mass resolution of the EPD SIS sensor. In Sect.~5, we use this solar particle event to discuss the consistency of EPT, SIS, and HET ion measurements for events that show high abundances of heavy elements.

A series of ion {$^3$}He-rich solar particle events that occurred between 17 and 23 November 2020, as shown in (e), were investigated by \citet{Bucik_21}. They used high-resolution STEREO-A imaging observations to locate the solar sources at two active regions near the east limb and find that these sources are well connected to Solar Orbiter's nominal footpoint. 

\citet{kollhoff-etal-2021} studied the large event at the end of November 2020, shown in (f), from a multi-spacecraft perspective, but focusing on Solar Orbiter/EPD observations. This is the first widespread event observed during the beginning of solar cycle 25 and it exhibited relativistic electrons and protons above 50 MeV. \citet{Mason-etal-2021b} analyze the heavy ion composition of this very same event, shown in (g), using ACE, STEREO-A, PSP, and Solar Orbiter measurements. They find that its properties coincide with those found for events observed at 1 au and that the spectra could be fitted by broken power law functions.

Around 10-11 December, another major event was measured for electrons and ions, indicated by (h), where the observed ions reach energies up to several tens of MeV, resulting in a significant intensity increase in the HET sensor. In Sect.~3, we therefore use this event, which could be observed with all four EPD sensors, to illustrate the measurement capabilities of EPD and to compare the spectra of the different sensors.

Other interesting periods in which EPD measurements have contributed to the understanding of the radiation conditions include: the encounter with the tail of comet C/2019 Y4 (ATLAS), labeled (I), as well as the first Venus Gravity Assist Maneuver (GAM) of Solar Orbiter, indicated by (II) in Fig.~\ref{fig:orbits}. 
The former is studied by \citet{MatteiniEtal_21}, who are able to identify Solar Orbiter's crossing of the ion tail of comet ATLAS, near 0.5 au. They found clear signatures of magnetic field draping around a low field region, which they interpret as the magnetotail structure, together with a local increase in the ion flux observed by STEP. The Venus GAM, occurring in late December 2020, provided interesting information about this planet’s induced magnetosphere. \citet{Allen2021} used observations from EPD and other instruments on Solar Orbiter to analyze the acceleration processes of suprathermal and energetic particles along Venus’ remarkably long magnetotail as well as the decrease observed in galactic cosmic rays (GCRs) during the closest approach.

\begin{figure*}[h!t]
    \centering
    \includegraphics[width=0.93\textwidth]{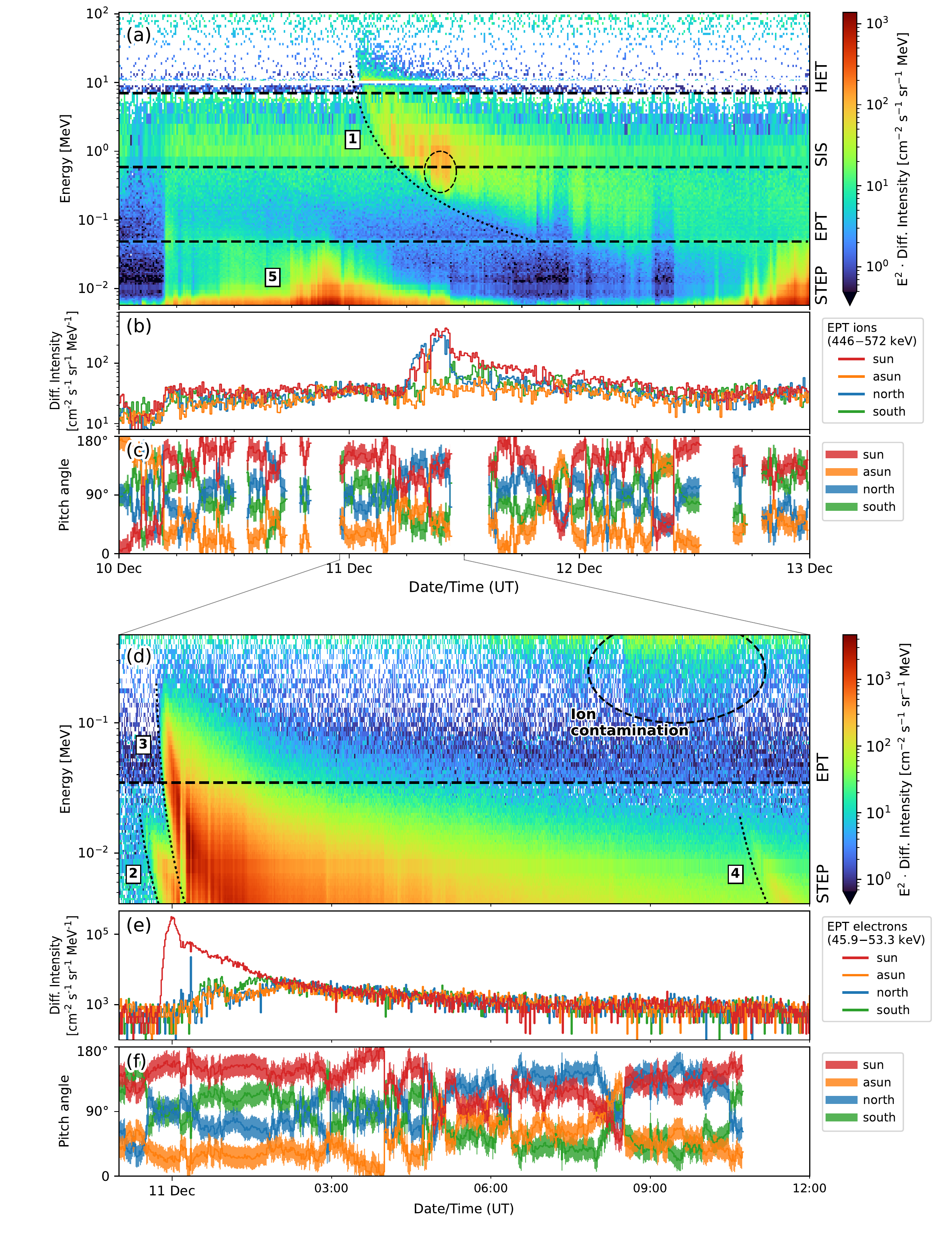}
    \caption{Dynamic spectra, sectored intensity time profiles, and pitch-angle coverage of EPT for the events seen on 10-11 December 2020 for ions (upper part) and electrons (bottom part). SIS and HET measurements in the upper part are for protons only, while STEP and EPT show data for all ions. The activity starting around 11:00 on 11 December (4) and the one visible in the low-energy STEP ion data that precedes the event on 10 December (5), are not associated with this solar event. Ion contamination of the electron measurements appears at high energies around 06:00 on 11 December 2020 and it is highlighted surrounded by an ellipsoid in panel (d); the ion population causing this signal is equally indicated in panel (a).}
    \label{fig:dyn_spectra}
\end{figure*}

\begin{figure*}[ht!]
    \centering
    \includegraphics[width=\textwidth]{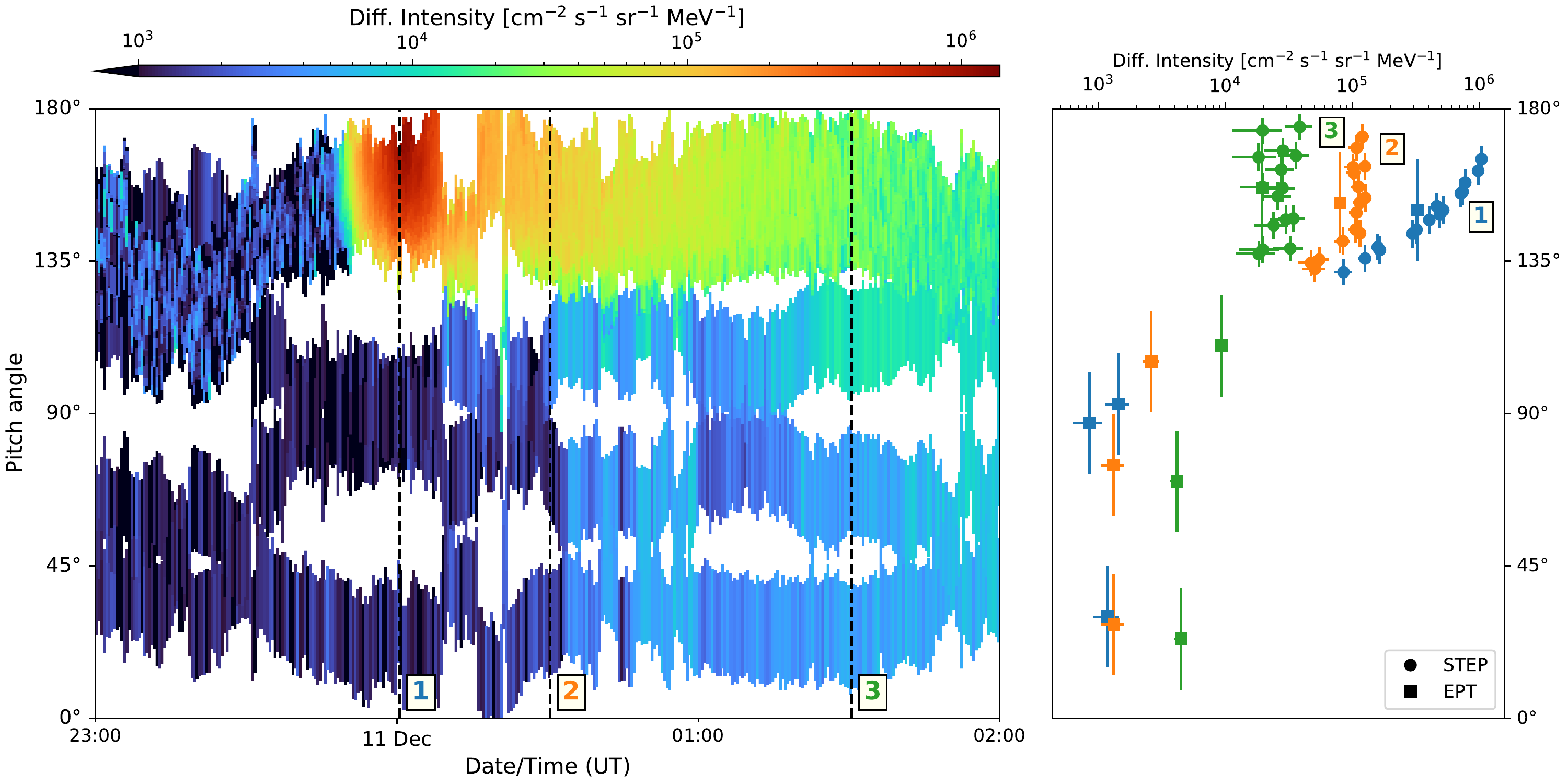}
    \caption{Reconstruction of the electron pitch angle distribution for the solar particle event on 10-11 December 2020 in the energy range from 30 to 50 keV using data from STEP and EPT. The panel on the right shows three slices (indicated by 1, 2, and 3) through the pitch-angle distribution measured by STEP (filled circles) and EPT (filled squares) at three different times in the event as indicated by the corresponding labels (1, 2, and 3) in the bottom of the left-hand panel. We can clearly see how the particle intensity changes from a beam-like distribution to a more isotropic one as time progresses. }
    \label{fig:pad}
\end{figure*}

\begin{figure}[ht!]
    \centering
    \includegraphics[width=\columnwidth]{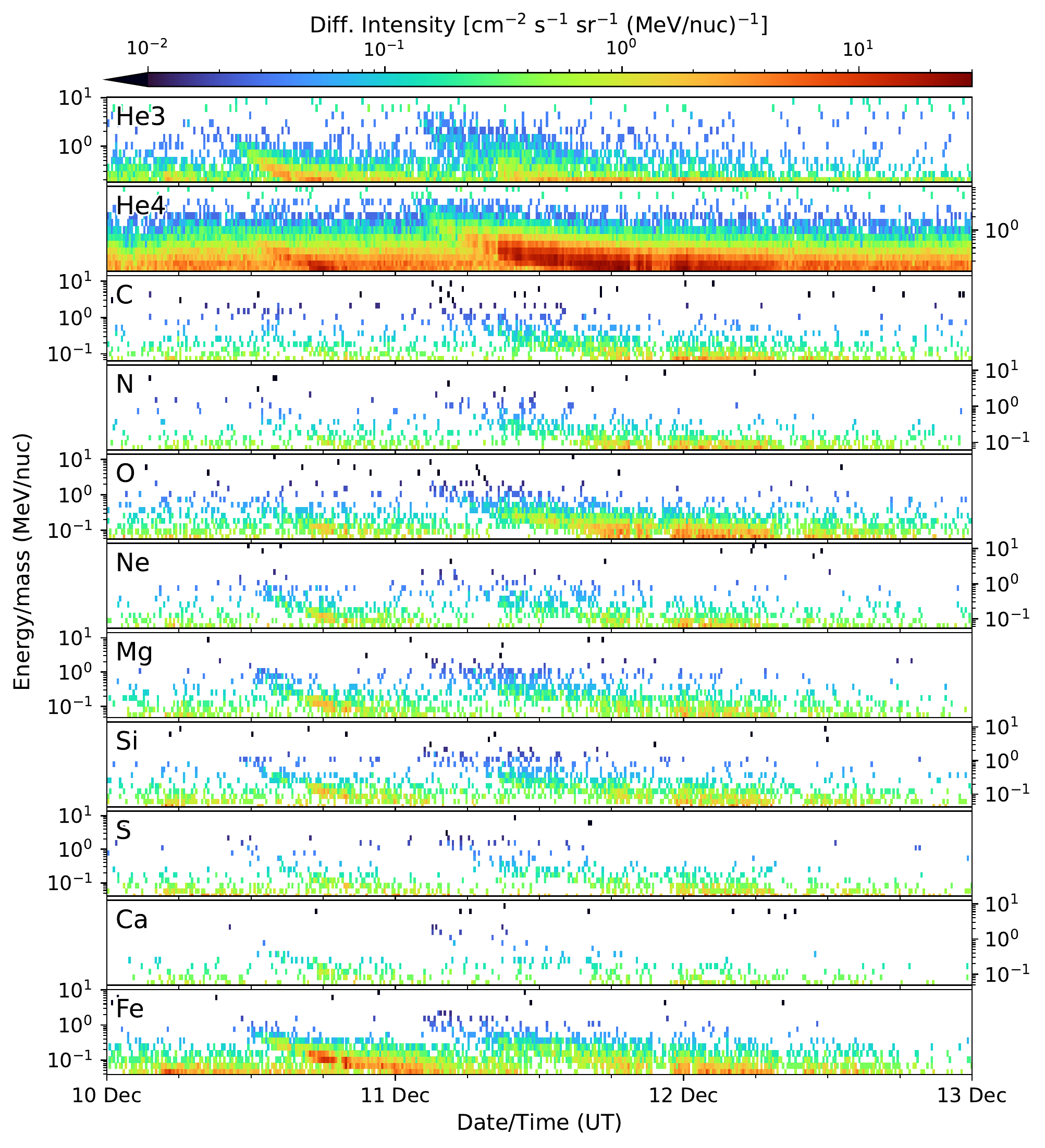}
    \caption{Dynamic spectra of the events on 10-11 December 2020 as seen by the SIS sunward telescope for the ion species included in SIS level 2 data (excluding hydrogen, which is already shown in Figure \ref{fig:dyn_spectra}).}
    \label{fig:sis_dyn_spectra}
\end{figure}

\section{10-11 December solar particle event}
\label{sec:Dec_event}

In this section we discuss, in greater detail, the 10-11 December 2020 event, seen in the right-hand panel of Fig.~\ref{fig:orbits}.
The electrons were first observed just before midnight on 10 December, high-energy ions (with energies of about 40\,MeV) followed a little later, just before 01:00 on 11 December 2020.

Figure~\ref{fig:dyn_spectra} shows a composite of data products from the different sensors of EPD for the 10-11 December 2020 solar event when Solar Orbiter was about 0.82 au away from the Sun. The upper half shows ion data at a time resolution of five minutes: (a) the dynamic spectrum for the sunward-pointing sensors, (b) intensity time profiles of all four EPT telescopes in the range of ion energies between 446\,--\,572 keV, and (c) the pitch-angle coverage of EPT\footnote{We make use of the nominal science data from the the public Solar Orbiter Archive (SOAR) \url{http://soar.esac.esa.int/soar/}}. The lower half shows electron data for a shorter time period at one minute resolution. The dynamic electron spectrum measured in the sunward-pointing telescopes is shown in (d), intensity time profiles for electron energies between 45.9\,--\,53.3 keV for the four EPT telescopes in (e), and the EPT pitch-angle coverage in (f).
 The sensors that acquired the data are indicated on the right-hand margin of the plot. To enhance the visibility of the velocity dispersion, the dynamic spectra have been scaled by $E^2$. STEP electron data were obtained by subtracting the STEP ion data (magnet channel) from the integral channel and applying the corresponding electron energy bins.
 
 The velocity dispersion of ions and electrons is clearly seen in their dynamic spectra. The onset of ions (labeled (1) in the SIS energy range) is seen at energies that trigger HET's BGO\footnote{BGO: Bismuth Germanate (Bi$_4$Ge$_3$O$_{12}$)} scintillator detector and corresponds to protons with up to $\sim 50$ MeV primary energy. At lower energies, protons seen by SIS also exhibit velocity dispersion. At even lower energies, EPT and STEP data continue to show velocity dispersion. Because these sensors cannot discriminate between different ion species in this energy range, we can only infer that they observe primarily protons from measurements by SIS. At low energies, STEP also shows a high ion intensity between about 05:00 on 10 December (5) and about 11:00 on 11 December, which is unrelated with the event discussed here. This ``background'' is real and is a consequence of the instantaneous direction of the interplanetary magnetic field (IMF) and an increase in suprathermal particles which is also seen in EPT\footnote{The solar wind velocity can also have an appreciable effect on STEP measurements through the Compton-Getting effect \citep{compton-getting-1935, ipavich-1974}.}. Panel (b) shows the strong anisotropy of 446\,--\,572 keV ions (in the space-craft frame), which is persistent during a large part of the event. The prompt increase in intensity is mainly observed by the sunward and northward pointing telescopes with some strong fluctuations in all EPT fields of view (FOVs). We can observe, for example, a sharp dent in the intensities between 08:00\,--\,09:00 on 11 December. These intensity variations are caused by local fluctuations in the IMF as can be seen in the sudden changes in the pitch-angle coverage of the FOVs shown in panel (c) of Fig.~\ref{fig:dyn_spectra}.
The data gaps in the pitch-angle coverage are due to EMC disturbances generated by the spacecraft or the payload. In such situations, the MAG data are not deemed to be of sufficient quality to be included in the SOAR.

In the lower half of Fig.~\ref{fig:dyn_spectra}, panel (d) shows that there were two separate electron injections (2 and 3) associated with this event. The lower-energy electrons (2) appear as the IMF changed its direction into the STEP FOV around 23:30 on 10 December 2020, the higher-energy electrons follow later (3), around 23:45. The low-energy injection is not unusual, a similar one is visible around 10:45 on 11 December (4). In fact, STEP sees two more such electron injections later on in the event discussed here (not shown). On the other hand, the increase of very high energy electrons around 10:00 am on 11 December 2020 (circled area annotated by ``ion contamination'') is not real, but rather the effect of ions (also circled in panel (a)) that penetrate the polyimide layer of the EPT electron telescope, which becomes clear when comparing the detailed temporal behavior of particles in the EPT ion channel, as well as the protons measured by SIS at that time. These ions have enough energy to penetrate the polyimide layer; but they stop in the electron detector and, thus, they are not measured in the adjacent ion detector, which would trigger an anti-coincidence veto. The velocity dispersion of these ``fake electrons'' precisely follows the velocity dispersion of the protons and ions. This serves to underline the need to compare electron data with data from the ion telescopes before interpreting EPT (and STEP) electron data (see Sect.~\ref{sec:in_flight_calib} for further details). We observe even stronger anisotropies for electrons than for ions as is shown in the intensity time profiles of 45.9\,--\,53.3~keV electrons in panel (e). The onset of the event is observed only by the Sun telescope while the remaining three only show a delayed gradual increase around 30~min later, but lasting several hours. The spike seen in the north telescope early on 11 December is real and due to a short-term fluctuation in the direction of the IMF and the ensuing change in pitch-angle coverage as can be seen in panel (f).

The two bi-directional or double-ended EPT-HET sensors provide four viewing directions that are fixed in the spacecraft frame, as are the viewing directions of the 15 STEP pixels (their pointing directions are given in the instrument paper). As the IMF is swept across Solar Orbiter, the instantaneous pitch angles accessible to EPD vary with time as illustrated in Fig.~\ref{fig:pad}, again using electron data for the 10-11 December 2020 event\footnote{Data are plotted at a time resolution of 30 seconds, using STEP energy bin 6 and EPT bins 0 -- 3 from the EPD data in the SOAR.}. During the rising phase of the event, most of the electrons arrive almost aligned with the magnetic field vector (see also sector intensities in the bottom panel (d) of Fig. \ref{fig:dyn_spectra}). The intensities progressively turn more isotropic during the decay phase of the event. This behavior is often observed during SEP events as a consequence of interplanetary scattering. The evolution of the pitch-angle distribution (PAD) is shown more clearly by the snapshots shown at three different times (1, 2, and 3) in the right-hand panel of Fig.~\ref{fig:pad}. In this representation the pitch angle is shown along the $y$-axis and differential intensity along the $x$-axis to be consistent with the neighboring left-hand panel. Just how narrow this electron beam is early on in the event becomes apparent when considering the pitch angles covered by the 15 STEP pixels (shown in the right-hand panel of Fig.~\ref{fig:pad}). The intensity at a pitch angle of $\sim 165^\circ$ is about ten times higher than that at $\sim 135^\circ$ which corresponds to a $1/e$ width of only $\sim 13$ degrees. Such high-resolution pitch-angle data from STEP will be extremely valuable when investigating microscopic processes such as pitch-angle scattering in similar events. It will also be helpful when analyzing and interpreting the PADs measured by EPT. By construction, EPT cannot distinguish the incoming directions of the particles arriving through one of its aperture cones. Therefore, all particles measured by an individual telescope are assigned to the same pitch angle, namely, the central pointing direction of that telescope. During periods of very anisotropic intensity, this can result in an over- or underestimation of the intensity measured at this central pitch-angle. STEP is able to provide a better resolution of the actual PAD inside its single field of view allowing us to assess this effect, at least for EPT observations of the sunward telescope.

EPD also provides information about the composition of the suprathermal ion population\footnote{The HET also provides composition information, but at higher energies. Such were not observed in this event, but see \citet{mason-etal-2021} for an example of HET observations during quiet times.}, as is shown in Fig.~\ref{fig:sis_dyn_spectra}, which shows dynamic spectra during the same time period as Fig.~\ref{fig:dyn_spectra} but for a number of different ion species measured with SIS, from $^3$He to Fe. Intriguingly, a preceding ion injection is seen which is not or only barely visible in protons beneath the label (1) in Fig.~\ref{fig:dyn_spectra}. From the data shown in Fig.~\ref{fig:sis_dyn_spectra}, it is clear that there are strong enhancements of $^3$He, $^4$He, and Fe in the 10 December injection but less so in C and N. Clearly, this event requires careful analysis, especially in view of a likely mass-dependent fractionation and $^3$He enrichment. These details illustrate that the Sun exhibited a high level of activity around the time of this event and that unraveling the dynamic spectra from individual injections will provide a rich source of information.

\section{In-flight calibration and operation of the EPD sensors}
\label{sec:in_flight_calib}

\begin{figure}[ht!]
    \centering
    \includegraphics[width=\columnwidth]{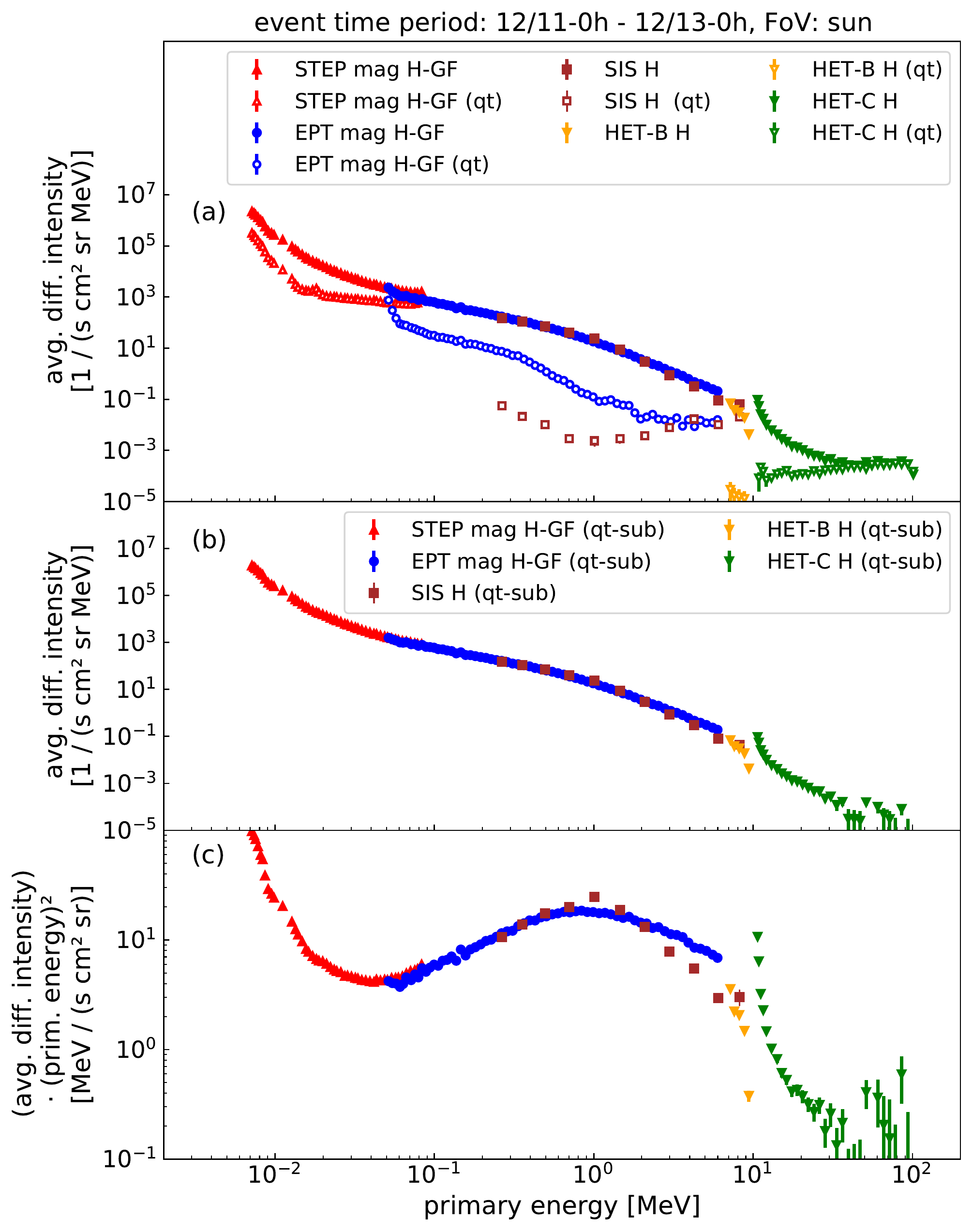}
    \caption{Average ion intensity spectrum of the 11-13 December 2020 solar particle event measured with all sunward-facing sensors of the EPD suite. The shown ion data products of SIS and HET contain only protons while for STEP and EPT also heavy ion species contribute to the measured intensity in the magnet channels with an estimated relative contribution on the order of 10\% for this event (see text for details). Panel (a) shows the measured ion intensity spectrum averaged over the two-day event period of 11-13 December represented by the filled symbols and also a quiet-time spectrum averaged over 7-12 November marked with the empty symbols (and labeled as "qt"). This quiet-time spectrum is subtracted from the measured spectrum to derive the quiet-time-corrected event spectrum in panel (b), labeled as "qt-sub". 
    In panel (c), the quiet-time-corrected spectra are multiplied with the square of the primary energy for a better visualization of the differences between the different sensor spectra. The differences between the EPT and SIS measurements, as well as the ones between the HET-B and HET-C protons, are discussed in the text.}
    \label{fig:December_spectra}
\end{figure}

While the various EPD sensors were calibrated before launch, not all aspects could be tested. For instance, geometry factors could not be entirely calibrated in the laboratory before launch. In Fig.~\ref{fig:December_spectra}, we present a combined ion spectrum measured with the whole EPD suite for the 10-11 December event, for which we observe systematically increased ion intensities from the beginning of 11 December up to roughly the beginning of 13 December. We show the average ion intensity spectra during the indicated time period for STEP ions in the magnet channels ('STEP mag H-GF'), EPT ions in the nominal proton magnet channels (`EPT mag, H-GF'), SIS protons ('SIS H'), HET protons stopping in detector B ('HET-B H'), and HET protons stopping in detector C ('HET-C H')\footnote{ We utilized level 2 data from the SOAR. In detail: STEP rates, EPT sunward telescope rates (5\,s cadence), SIS sunward ("a") telescope medium rates and HET sunward rates. As this is not explicitly labeled in the SOAR data, we point out that the first five energy bins of the HET data correspond to particles that stop in detector~B.}. These spectra shown in panel (a), and all marked by the filled symbols, have not been corrected for instrumental background. STEP and EPT are more affected by instrumental background than the other sensors because they rely on single detector counts without any anti-coincidence in the case of STEP and on single detector counts with the adjacent detector used as anti-coincidence in the case of EPT. In panel (a), we also show a second set of data points marked with empty symbols for each sensor which were acquired over the five-day period from 7-12 November to serve as an estimate for the quiet-time background (labeled "(qt)"). The background correction is important because the December event did not show very high intensities (despite being one of the bigger events observed so far with Solar Orbiter).

In the middle panel of Fig.~\ref{fig:December_spectra}, we present the composite spectrum for the 11 December event measured with all EPD sensors, but now with the quiet-time spectrum subtracted. 
This reduces the background in STEP and EPT that results, for instance, from GCRs that penetrate the detectors and deposit low-energy signals, but also from the permanent quiet-time anomalous cosmic rays (ACR) and GCR populations measured at high energies by SIS and HET. 
In the lowest panel, the background-corrected spectra are multiplied with the square of the primary particle energy, $E^2_{\rm prim.}$. This representation allows a clearer illustration of the differences between the spectra measured by the individual sensors because it compensates somewhat for the slope of the quiet-time-corrected spectrum.

The composite event spectrum measured with the different EPD sensors (b) shows remarkable agreement across more than three orders of magnitude in energy and ten orders of magnitude in particle intensity. However, there are two obvious shortcomings, which become much more apparent in panel (c). First, measurements of protons by SIS and ions in the magnet channel of EPT begin to disagree at around 1\,MeV. To a certain extent, this is expected as both STEP and EPT only measure total energy and, thus, cannot discriminate between different elements. On the other hand, SIS, as a time-of-flight versus total energy instrument is capable of discriminating even the isotopes of many elements. For the particle event discussed here, this effect only plays a minor role because the EPT magnet channels are dominated by protons. This is found by comparing with SIS heavy ion measurements, but not shown here (We will discuss the importance of heavy ions in more detail in Sect.~\ref{subsec:EPDsensor_comparison}). Nevertheless, the heavy ions are not completely negligible, as they contribute about 10\% to the intensity in the overlapping energy range between EPT and SIS.

The relatively smooth transition between STEP and EPT in the event spectrum makes it reasonable to assume a similar heavy ion fraction in the STEP magnet channels as in EPT. For the sake of completeness, we also mention that the ion intensities in STEP and EPT (in the nominal proton energy range below 6.1 MeV) are always calculated as if all particles were protons (marked by "H-GF" in Figure \ref{fig:December_spectra}), namely, by using the proton geometry factors, efficiencies and energy loss in the respective SSD dead layer.
 In the nominal He energy range for EPT, the intensity is calculated analogously by assuming the specific quantities for $^4$He. We also note that the proton detection efficiencies for the two highest energy channels of SIS (above~10\,MeV) could not be calibrated in flight thus far, due to the low count statistics for the observed events; thus, these data points are left out in Fig. \ref{fig:December_spectra}. A combined analysis of EPT and SIS spectra is given in Sect.~\ref{subsec:EPDsensor_comparison}.
 
A second, considerably larger discrepancy is seen around 10 MeV. Protons of this energy penetrate HET's A and B detectors and may just have enough energy to trigger HET's BGO scintillator which is labeled as detector C \citep[see][for details]{elftmann-2020}. This is wrapped in several layers of highly reflective millipore and teflon to ensure a maximum retention of the scintillation light released by the BGO, but the exact behavior of HET's response function at this threshold energy could not be calibrated in the lab. We have not yet found a quantitative explanation for this obvious inconsistency between the two detectors, but interestingly, the $^4$He spectra accumulated during quiet time periods do not exhibit this discrepancy \citep{mason-etal-2021}. We are therefore eagerly awaiting a large, high-energy particle event, which will allow us to unambiguously calibrate the overlap between EPT, SIS, and HET.

\subsection{STEP}
\label{subsec:calib_STEP}
Since its commissioning at the end of February 2020, STEP has been operating nominally at almost constant temperature and as expected. STEP has two sensor heads, which function similarly to a pinhole camera. Particles from all directions in STEP's FOV enter through the pinhole and are subsequently measured in a 3 $\times$ 5 multi-pixel solid state detector. One of the two sensor heads has a deflection magnet that deflects electrons, thus measuring only ions (and occasionally energetic neutral atoms). Fig.~\ref{fig:dyn_spectra} shows ion (upper panel, a) and electron (lower panel, c) measurements by STEP. The 15 detector pixels allow for very fine pitch-angle resolution, as illustrated in the upper right corner of Fig.~\ref{fig:pad}. 

Because electrons and ions of the same kinetic energy don't experience the same energy loss in the dead layer of the detector, the energy ranges for electrons and ions are not the same. After carefully adjusting the thresholds of STEP during the commissioning phase, the lowest energy bin for electrons (protons) now starts at 4.2 (5.7) keV. It is important to note that we give the lower threshold for protons here, but STEP cannot discriminate between different ion species. In the presence of suprathermal He and heavy ions these can contribute significantly to the STEP measurements. For more details, see the discussion in Sect.~\ref{subsec:calib_EPT}, which is also pertinent to STEP. Such contamination by heavy ions or suprathermal particles from the solar wind may be the reason for the increase in suprathermal ions seen in panel (a) of Fig.~\ref{fig:dyn_spectra}, which is well ahead of the time for ions from the event and in the STEP energy range to arrive. This can be easily recognized as they are present well before the dotted curve marking the proton-ion velocity dispersion. 
Heavy solar wind or suprathermal ions may also be contributing to the lowest energy bins seen in Fig.~\ref{fig:December_spectra}. The lowest energies covered by STEP are very close to those of heavy solar wind ions. This means that they can trigger STEP if they happen to enter STEP from the right direction. The problem is that at a given energy, the phase space volumes for different ion species are very different. At these low energies and close to the bulk solar wind, energy and directional information is mixed if transformed to the physical frame of reference.
\cite[See][for an in-depth discussion of the correct reference frame.]{nemecek-etal-2020} This also means that at these low energies, it is absolutely critical for pixel data to be analyzed to infer properties of the velocity spectrum of suprathermal ions. Of course, this effect has additional implications for the separation of electrons and ions because of the additional small but velocity-dependent position shift of the ions in the magnet channel. 

We are currently preparing an update to the STEP data products, which will be uploaded prior to the end of 2021 and aims to increase both the time and the energy resolution of the pixel-wise energy spectra. This will allow for better studies of kinetic processes such as wave-particle interactions, along with improved resolution in phase space, as discussed above. STEP telemetry also includes a very limited number of full-resolution pulse-height analysis (PHA) words, which has allowed us to define some minor adjustments to the energy calibration of individual pixels. These changes will also be uploaded in the near future. Using PHA data we found unphysical signatures at small energies in some of the pixels in the two sensor heads. This instrumental issue will be resolved by adding a small delay for reading out the signals in the application-specific integrated circuit (ASIC) at the cost of a slightly increased dead time.

To allow for an adaptation of the geometry factor, STEP can switch to a measurement mode that uses 15 smaller pixels that are placed next to the larger pixels in the sensor heads, respectively, with a sensitive area of 0.3\,mm$^2$ of each pixel. 
Because solar activity has been low in the past year, we have not yet had a chance to test the small pixels in the two sensor heads. As we expect that they will exhibit a different background than the large pixels, we will have to acquire background measurements for the small pixels in the near future. This means that STEP will only measure at a very much reduced count rate for that period of time. This issue will be appropriately flagged in the data made available in the SOAR.

\subsection{EPT}
\label{subsec:calib_EPT}
Since its commissioning at the end of February 2020, EPT is in nominal operation. The in-flight calibration performed during the first weeks of operations in March have led to a small change of the EPT energy range, which now starts at about 48 keV and ends at about 6.1 MeV for the nominal proton energy range. This is the result of corrected response model simulations for the sensor taking into account a more accurate description of the solid state detector (SSD) dead layers. The nominal He energy range is still the same as given in the instrument paper. EPT data available in the ESA SOAR use the thus corrected energy bins.

EPT makes use of the so called ``magnet-foil'' technique and is based on previous experience acquired during the design, construction, and operation of the STEREO/SEPT instrument \citep{Muller-Mellin_2008}. This technique allows us to accurately separate electrons from ions up to $\sim$425~keV.

In the magnet channels, only ions are measured with the instrumental limitation that no distinction between the ion species can be made because of the single energy measurement in EPT. Measurements of the composition of energetic ions are provided by SIS and HET. Thus, the contribution of heavy ions to the EPT (and STEP) ion measurements has to be estimated from SIS data in the overlapping energy range between EPT and SIS as discussed in Sect. \ref{subsec:EPDsensor_comparison}. 
A second issue, observed both in EPT and STEREO SEPT, is the contamination in the electron measurement by high-energy ions \citep{Wraase18a_SEPT,Wraase18b_SEPT} when such are present. EPT uses a foil at one of the entrances of each telescope to stop ions up to $\sim~400$~keV. However, this technique only provides a clean rejection of ions with energies low enough to stop in the front foil or high enough to reach the second detector (see Fig. 19 in the instrument paper). For that reason, electron measurements taken during periods of elevated intensity of high-energy ions, should be investigated carefully as they will almost certainly be affected by this contamination. An example is shown in the panel (c) of Fig.~\ref{fig:dyn_spectra}, where the ion contribution is clearly seen. The contamination in the electron channel appears in coincidence with the arrival of ions above 400 keV visible in panel (a). Periods affected by significant ion contamination of the electron channels can be identified by comparing the observations of the magnet and foil telescopes. For instance, during heavily contaminated periods, signatures of $\sim$50 keV ``electrons'' (foil) and $\sim$450 keV ``ions'' (magnet) follow almost identical dispersion curves and have very similar time profiles. Currently, we are working to develop an inversion procedure to efficiently separate this ion contribution.

Prior to April 2021, we plan to apply a change in the structure of EPT's data products. In order to make EPT data more user-friendly and scientifically interesting, we will implement a symmetrical energy binning for all magnet and foil channels, that is, the energy range, number of energy bins (77) and cadence for both detectors (close mode: 1 s; far mode: 5 s) will be identical. This will help in the interpretation of the data as the data products of both magnet and foil detectors will be directly comparable. It will also help with untangling the proton and helium contributions, because the energy bins will be extended across the range where protons penetrate the front detector and helium is then the predominant ion species. In turn, the high-cadence data products described in the instrument paper will be discontinued because we will then already be providing high energy resolution data at the highest cadence possible.
The time cadence in the far mode of EPT will be constant at 5~s for all data products. This simplification of the data products, one for ions and one for electrons, also eliminates the need of specific data products for burst modes, making the whole structure easier to handle. As mentioned before, we will also extend the energy range to higher energies for the level 1 data products. This will allow us to disentangle the different contributions of protons and heavier ions, as measurements will extend over the effective threshold of proton detection. The energy bins of the level 2 data products will remain unchanged.

\subsection{SIS}
\label{subsec:SIS}

Calculating particle intensities from SIS is straightforward since all ion species are clearly separated, and background is low, due to the time-of-flight versus energy coincidence. Detection efficiency for protons and He is less than 1.0, however, and needs to be determined in-flight since it depends on the energy loss of the particle in the foils, $\dd E/\dd x \cdot d_f$, and the gains of the micro-channel plate (MCP) stacks which in turn depend on the intrinsic gains of the plates as well as temperature and high-voltage (HV) bias. Because changes in efficiency may not be detected until after the data is analyzed, efficiency corrections are applied in ground-based processing, rather than on board.

The post-launch peak efficiency (near 300 keV/nucleon) is about 20\% for protons, 95\% for He, and 100\% for heavier ions \citep{RodriguezPacheco_2020}. Detailed efficiency curves for protons are obtained by comparing the SIS spectra with those from the EPT sensor for energies that overlap, and whose SSDs do not have significant efficiency uncertainties. Since EPT does not distinguish between ion species, this comparison is done in gradual SEP events that are dominated by protons. We used the 24-25 November 2020 event for this purpose, using the decay phase when anisotropies are small and therefore have little impact on the slightly different FOVs of SIS and EPT. Helium efficiencies are determined by the dual time-of-flight measurement in SIS, by measuring the fraction of He events in each energy bin that trigger the Start-2 MCP stack when the Start-1/Stop is triggered.
The calibrated efficiencies are included in the calculation of the SIS level 2 intensities. 
As the mission progresses, these SIS-EPT spectra comparisons will be updated, and if the SIS efficiency changes due to plate aging or radiation damage, the MCP HV bias can be adjusted to mitigate such effects.

After some delays due to the COVID-19 impact on the spacecraft operations center, SIS low voltage was powered up on 2 April 2020. After allowing additional time for outgassing, the high voltage ramp-up began on 17 April, and was completed on 24 April, 75 days after launch. The instrument has operated continuously since then with short interruptions due to spacecraft operations (e.g., thruster firings). The instrument operating temperatures at heliocentric distances between 0.5\,--\,1.0 au have been in the range -10 $^\circ$C to -20 $^\circ$C, but much lower if powered off. The spacecraft heater set points were raised to keep the instrument warmer if powered off, and are now set at -24 $^\circ$C and -20 $^\circ$C for telescopes A and B.

The on-board particle identification lookup tables, which were based on pre-launch calculations and particle calibration, have been updated to more accurately locate the mass tracks in the time-of-flight versus energy matrices. Even though the immediate post-launch period was during solar minimum there was luckily a \textsuperscript{3}He-rich SEP event on 21-23 July, which was intense enough to populate the H, He, and heavy ion tracks up to \textasciitilde1 MeV/nucleon \citep{Mason_2020}. An analysis of this event allowed for the generation of improved lookup tables, which were uploaded on 20 October 2020. These new tables also included small adjustments to compensate for the lower operating temperature of SIS, whose prelaunch particle calibration runs were done at room temperature.

Two further changes in SIS operations are anticipated. Firstly, the time-of-flight versus energy particle tracks have not been populated in the range between$~$1\,--\,20 MeV/nucleon, which requires an SEP or shock event more intense than any observed so far. After such events occur, it will be possible to make small adjustments to accurately center the higher energy mass tracks and a new set of lookup tables will be uploaded. Secondly, the SIS telescopes each have an independently operated mechanical door that can be closed in steps to reduce the geometry factor to 25\%, 5\%, or 1\% of fully open. The reduction of geometry factor is intended to allow SIS to return calibrated data in periods that would otherwise saturate the instrument due either to excessive currents in the MCPs or overloading the on-board processor. An on-board algorithm is required to detect such conditions and adjust the doors accordingly as was done on the ACE/ULEIS instrument \citep{Mason_1998}. When intense events occur, the SIS response will be analyzed to adjust the door closing parameters that detect saturation or overloading, reduce the door opening as required, and when intensities fall increase the door opening. The derived parameters will then be uploaded and the processor set to control the doors automatically. Until that time, the doors are controlled by ground command only.

\subsection{HET}
The High-Energy Telescope (HET) measures protons and heavy ions in the energy range from $\sim$7 MeV/nuc to a few hundred MeV/nuc (species-dependent) and electrons from 0.3 - 30 MeV. The two double-ended HET units are co-located with the two EPT units and share the same electronics boxes and field of view directions. HET uses the $\dd E/\dd x$ vs.\,total $E$ measurement principle to discriminate between different particle species and isotopes such as $^3$He and $^4$He. A viewing direction consists of two Si solid state detectors A (front) and B. The B detectors of the two opposing viewing directions sandwich a BGO scintillator (detector C). The B detector of the opposing viewing direction provides the anti-coincidence to discriminate against penetrating particles.

HET has been acquiring data since 28 February 2020, directly following the commissioning of EPD. On 12 May 2020, a minimum $\textrm{(A+B)}/\textrm{A}$ ratio (which is the sum of energy deposition in detector A and B divided by the energy deposition in detector A) of 1.2 was introduced in the ABnC (i.e., particles stopping in detector B) level 3 triggers of both HET units \citep[see Fig. 35 in][]{RodriguezPacheco_2020}. Earlier data showed that the lowest energy bins of the proton and $^4$He channels in this coincidence were contaminated by particles with $\textrm{(A+B)}/\textrm{A} < 1.2$. By utilizing additional Geant4 simulations, it was possible to confirm that the contamination was caused by secondary particles generated by heavy ions. While these changes of the configuration have resulted in reasonable intensities in the lower energy range of ABnC, as shown in Fig.~\ref{fig:December_spectra} for protons, there are still some issues in the HET data -- which we point out here.

As already mentioned at the beginning of Sec. \ref{sec:in_flight_calib}, we can see that the last data point of B (shown in yellow in Fig.~\ref{fig:December_spectra}) is too low and the first several data points of C (in green) are too high. A similar behavior is observed for electrons, but not for other ions. Furthermore, a comparison of the quiet-time GCR proton spectrum between 10\,MeV and 100\,MeV with that of EPHIN on SOHO \citep{mueller-mellin-etal-1995} shows the former is a factor of $\sim 2$ higher than the latter, while the helium spectra of both instruments are in agreement. The reason behind these broad discrepancies are being investigated by the EPD team. 
The HET spectra of $^4$He and heavier ions acquired during quiet times in EPD's first year of operation agree well with those measured by SIS and ACE/ULEIS at 1 au. The particle spectrum in this energy range is dominated by GCR and remnant solar particles. Some species, such as oxygen, also have a strong contribution from ACRs that is clearly visible \citep{mason-etal-2021}.

A little-advertised data product of HET is the count rate registered in the BGO scintillator crystal. This single detector count rate without any coincidence condition is dominated by the penetrating GCR intensity, which is modulated by short-term disturbances such as the passage of ICMEs or planetary fly-bys. \citet{vforstner-etal-2021} discuss the timing and amplitude of the 3\% Forbush decrease measured when an ICME passed over Solar Orbiter on 19 April 2020, and \citet{Allen2021} discuss the 5\% drop in count rates due to the GCR ``shadow'' of Venus during the December 2020 Venus flyby. Such small signals can be measured with good statistics thanks to the large geometric factor for this data product of $\sim 100$ cm$^2$ sr and the ensuing high count rate.

\section{Combined analysis of EPD sensor data}
\label{subsec:EPDsensor_comparison}

\begin{figure}
    \centering
    \includegraphics[width=\columnwidth]{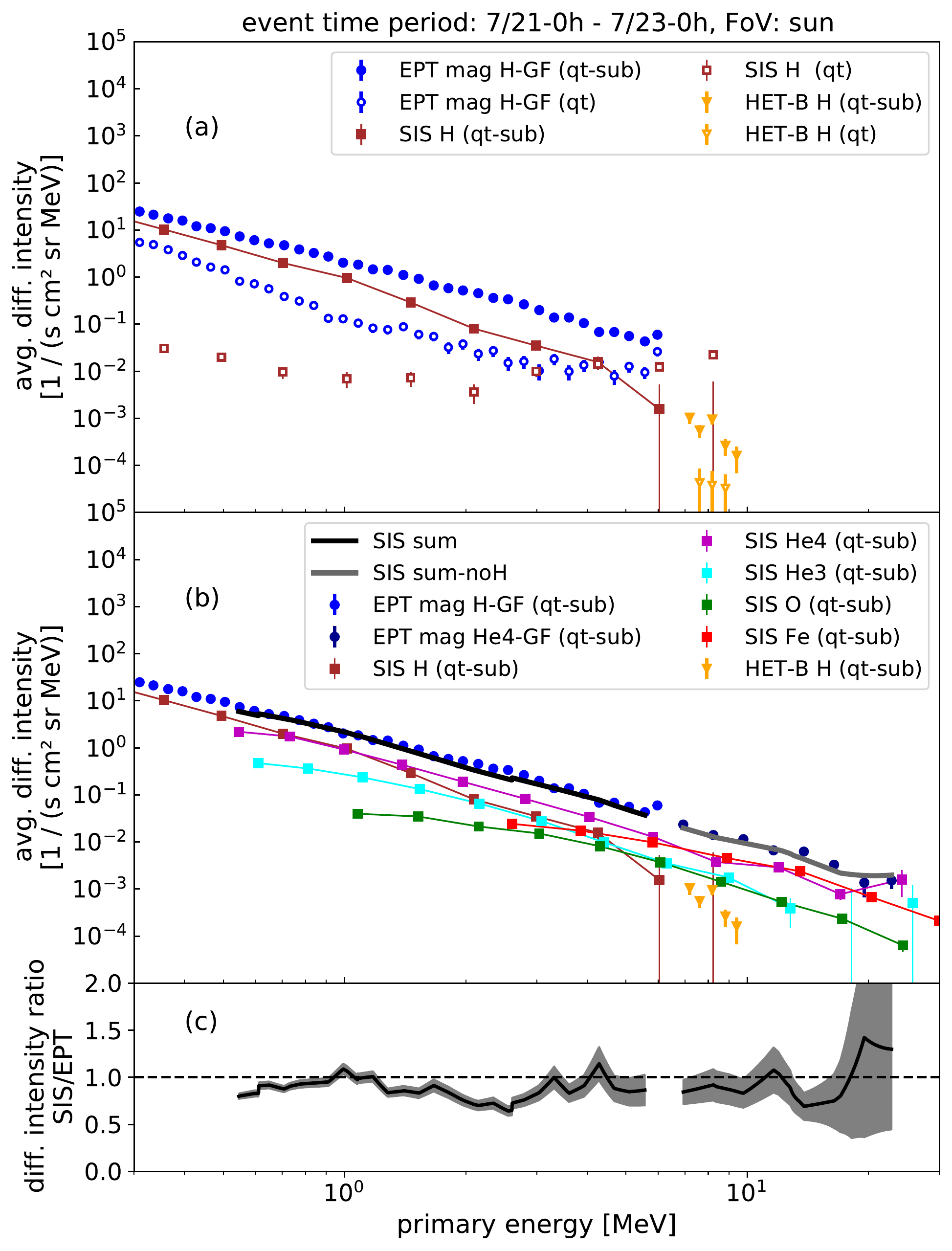}
    \caption{Comparison between EPT, SIS and HET data for the 21-23 July 2020 event for the sunward-facing telescopes.
    The upper panel shows the SIS and HET-B proton intensity together with the EPT ion spectrum in the magnet channel over the nominal H energy range. The filled symbols show the measured spectra after the subtraction of a quiet-time spectrum. This quiet-time spectrum was acquired between 18-20 July and is shown as empty symbols. Error bars are statistical. The noticeable increase of intensity in the quiet-time SIS proton spectrum above a few MeV is due to ACRs.
The middle panel adds the quiet-time-corrected SIS spectra for H, $^3$He, $^4$He, O, and Fe. The black line up to about 6\,MeV represents the sum of the SIS ion species intensities that can be compared to the EPT spectrum in the nominal H energy range as explained in the text. Above 7\,MeV we compare the EPT ion spectrum in the nominal He energy range and the sum of heavy ions measured by SIS as a gray solid line. 
 The bottom panel shows the ratio of the sum between the relevant SIS ion spectra and the EPT spectra to allow a better comparison. The gray-shaded area marks the propagated statistical errors.}
    \label{crosscalibration}
\end{figure}

The EPD instrument is designed as a suite of complementary sensors that are altogether capable of measuring a large set of particles over a wide energy range. Therefore, the most complete physical information can be obtained when the data of several of these sensors are analyzed together in the energy range of interest.
A good example for such a combined analysis is the utilization of SIS data to estimate the proton and heavy ion contributions in the particle spectrum measured by EPT. 
While the sensor discriminates between electrons and ions using the foil-magnet technique, it cannot discriminate between different ion species which stop in the front detector\footnote{The only exception is that we can exclude protons when the particles deposit more energy than is needed for protons to penetrate the front detector and trigger the second (anti-coincidence) detector. Therefore, EPT does have data products for particles which deposit energy in both SSDs of a given telescope.}. 
Because differential intensities of particles are often organized according to energy-per-mass (or particle speed), the importance of heavy ions becomes disproportionate when measured in total energy. This effect depends strongly on the spectral index of the event as can be understood from a simple example calculation:
let us assume that for the ions, the (differential) intensity in dependence on energy-per-mass follows a power law of:\

\begin{align}
\dfrac{dI_\alpha}{d(E/m_\alpha)}=I_0 \cdot \left( \dfrac{E/m_\alpha}{E_0/m_0} \right)^{-\gamma},
\label{Eg:EPT_spectrum_transform_1}
\end{align} 
where $\alpha$ denotes an ion species, $E/m_\alpha$ describes the ions' energy-per-mass, $\gamma$ is the spectral index of the power law, and $I_0, E_0$, $m_0$ are the necessary scaling parameters\footnote{We note that $I_0$ has the dimension of a differential intensity $dI/d(E/m)$ here.}. 
The energy-per-mass-dependent intensity can be related to the corresponding energy-dependent intensity that is measured with EPT by 
\begin{align}
\dfrac{dI_\alpha}{dE}=\dfrac{dI_\alpha}{d(E/m_\alpha)}\cdot \dfrac{1}{m_\alpha} \ .
\label{Eg:EPT_spectrum_transform_2}
\end{align} 
For the sake of simplicity, we now consider the case that two ion species with masses $m_a>m_b$ have exactly the same intensity $dI_a/d(E/m_a)=dI_b/d(E/m_b)$. Combining Eq. \ref{Eg:EPT_spectrum_transform_1} and Eq. \ref{Eg:EPT_spectrum_transform_2}, we find the following for the ratio of the measured energy-dependent intensities of the two species:
\begin{align}
\left.\left( \dfrac{dI_a}{dE} \right) \right/ \left( \dfrac{dI_b}{dE} \right)=\left( \dfrac{m_a}{m_b} \right)^{(\gamma-1)} \ .
\end{align} 
We can see that for spectral slopes $\gamma>1,$ the heavier species is over-represented in the EPT measurements. In particular, the measured heavy ion contamination in EPT is highest for events with a soft spectrum that are rich in heavy elements.
Such an event is illustrated in Fig. \ref{crosscalibration}\footnote{We utilized level 2 data from the SOAR. In detail: STEP rates, EPT sunward telescope rates (regular 5\,s cadence), SIS sunward ("a") telescope medium rates and HET sunward rates.}. The upper panel shows the already quiet-time-corrected spectrum for the 21-23 July 2020 event marked by the filled symbols (and labeled "qt-sub"). The subtracted quiet-time spectrum is also shown as empty symbols (and labeled "qt"). The ions measured by EPT in the (nominal H) magnet channel are depicted in blue. Protons measured with SIS (HET) are shown in brown (yellow). 

In this representation, the EPT ion and the HET proton differential intensities appear to deviate by almost two orders of magnitude while the SIS proton spectrum lines up nicely with the HET protons. The difference between the EPT ion spectrum and the SIS proton spectrum is due to the disproportionate contribution of heavy ions discussed above. This particle event was rich in heavy ions, which makes it an ideal candidate to illustrate this effect in the middle panel of Fig. \ref{crosscalibration}. There we also show SIS $^{3}$He, $^{4}$He, O, and Fe spectra with their pre-event background again subtracted. Adding all these ions to the protons measured by SIS results in the black curve which agrees well with the EPT measurements\footnote{The comparison does not include the highest EPT nominal H energy channel, as the intensity calculation is off for this channel. This is due to an underestimated geometry factor for this channel which will be corrected in a future EPT data update.}. 
 
We also show (in filled dark blue circles) the quiet-time-corrected spectrum measured by EPT between 7 and 25 MeV in the magnet channel. Because protons penetrate the detector at these energies, these measurements have to be due to He or heavier ions. The intensities in this nominal "He4" range are calculated with the EPT helium response, that is slightly different from those of protons for the lower energy range. 
We again compare this EPT spectrum with the sum of all heavy ion species (with $Z\geq2$) measured with SIS shown as a solid gray line. The agreement is again much better than if we had only considered $^4$He. 
The bottom panel of Fig. \ref{crosscalibration} shows the ratio of the sums of the individual ion spectra measured by SIS to the ion spectrum measured by EPT (solid line) as well as the propagated statistical error of this quantity (shaded).

Thus, the contribution of protons and heavy ions to the ion spectrum measured by EPT can be quantified as follows:
\begin{itemize}
\item Subtract an appropriate quiet-time period spectrum, as explained in Section \ref{sec:in_flight_calib}. This removes the instrumental background as well as the ubiquitous GCR and ACR population from both EPT and SIS data.
\item Express the ion intensities measured by SIS in terms of energy rather than energy-per-nucleon to allow comparison with the EPT spectrum (see upper and middle panel of Fig. \ref{crosscalibration}).
\item Subtract the sum of the heavy ion ($^3$He - Fe) intensities measured by SIS from the EPT ion intensities at each energy. This provides an estimate for the proton intensity contributing to the EPT measurement.
\item Finally, compare the sum of all ion species measured by SIS (including SIS protons) to the EPT ion intensity to get a conservative estimate of the uncertainties for the derived proton fraction in the EPT spectrum (see middle and bottom panel of Fig.~\ref{crosscalibration}).
\end{itemize}

This procedure results in a proton spectrum that is derived from a set of consistent measurements at the time of the event and makes use of the actual abundances and spectral slopes of the different ion species as measured by SIS. This approach is especially robust because it is independent of averaged information on composition, which is known to vary from event to event and from particle source to particle source \citep{Desai_2006}.

The method described above will require some refinements, which we are currently working on and which will be presented elsewhere. For example, the particle intensity in EPT is calculated using proton geometry factors for all ions measured in the magnet channel; moreover, the energy loss of very heavy ions in the entrance windows of the EPT detectors as well as the pulse height defect in the SSDs have not been accounted for. To include these effects we are currently developing a more detailed response model for EPT that accounts for the influence of heavy ions across all of EPT's FOV. These simulations can be validated using SIS measurements of heavy ions if the different FOVs and other instrumental properties are accounted for. The excellent mass resolution of the SIS instrument in combination with two time-of-flight systems allows very accurate determination of particle mass and energy with high detection efficiencies for all heavy ion species.
Yet, as the SIS sensor is designed for these heavy ion measurements, the SIS micro-channel plate gains are set to have relatively low H efficiency in order to reduce instrument triggering rates in large SEP events so that the heavy ion intensities can be accurately measured\footnote{This same approach was used on ACE/ULEIS and STEREO/IMPACT/SIT (see \cite{Mason_1998} and \cite{Mason_2008}).}, while proton intensities are handled by EPT. This has the consequence that the SIS detection efficiencies for protons are significantly lower than for the heavy species and are subject to further refinement during the ongoing in-flight calibration. This refinement is done based on data from measured events with a dominant proton component compared to all heavy species, such as the 24-25 November event (see Section \ref{subsec:SIS}).

Despite these caveats, we can see from Figs.~\ref{fig:December_spectra} and \ref{crosscalibration} that the agreement between EPT and SIS spectra is already remarkably good given the fact that we have still not observed a large particle event which would populate spectra across the entire SIS energy range. We expect the observed small differences to decrease with progress in the ongoing in-flight calibration of EPT and SIS.

\section{Possible effects of spacecraft operations on the data}
\label{sec:ops}

Spacecraft operations can affect EPD's measurements and have done so to various degrees on some occasions. In the worst case, EPD is switched off which - obviously - results in a total loss of data for the corresponding time period. Other, more benign examples include the firing of thrusters, spacecraft rolls, or slews. 

Such operations are typically planned months in advance, although they are sometimes performed in a responsive way (e.g., triggered by temperature conditions) and can only be labeled a-posteriori from the telemetry rather from the planning products.

In this section, we describe the most important maneuvers affecting the instrument data and present (in App.~\ref{app:ops}) a list of them for the first year of EPD operations. Data that are adversely affected by spacecraft operations are flagged accordingly in the SOAR.

\begin{figure}[h!t]
    \centering
    \includegraphics[width=\columnwidth]{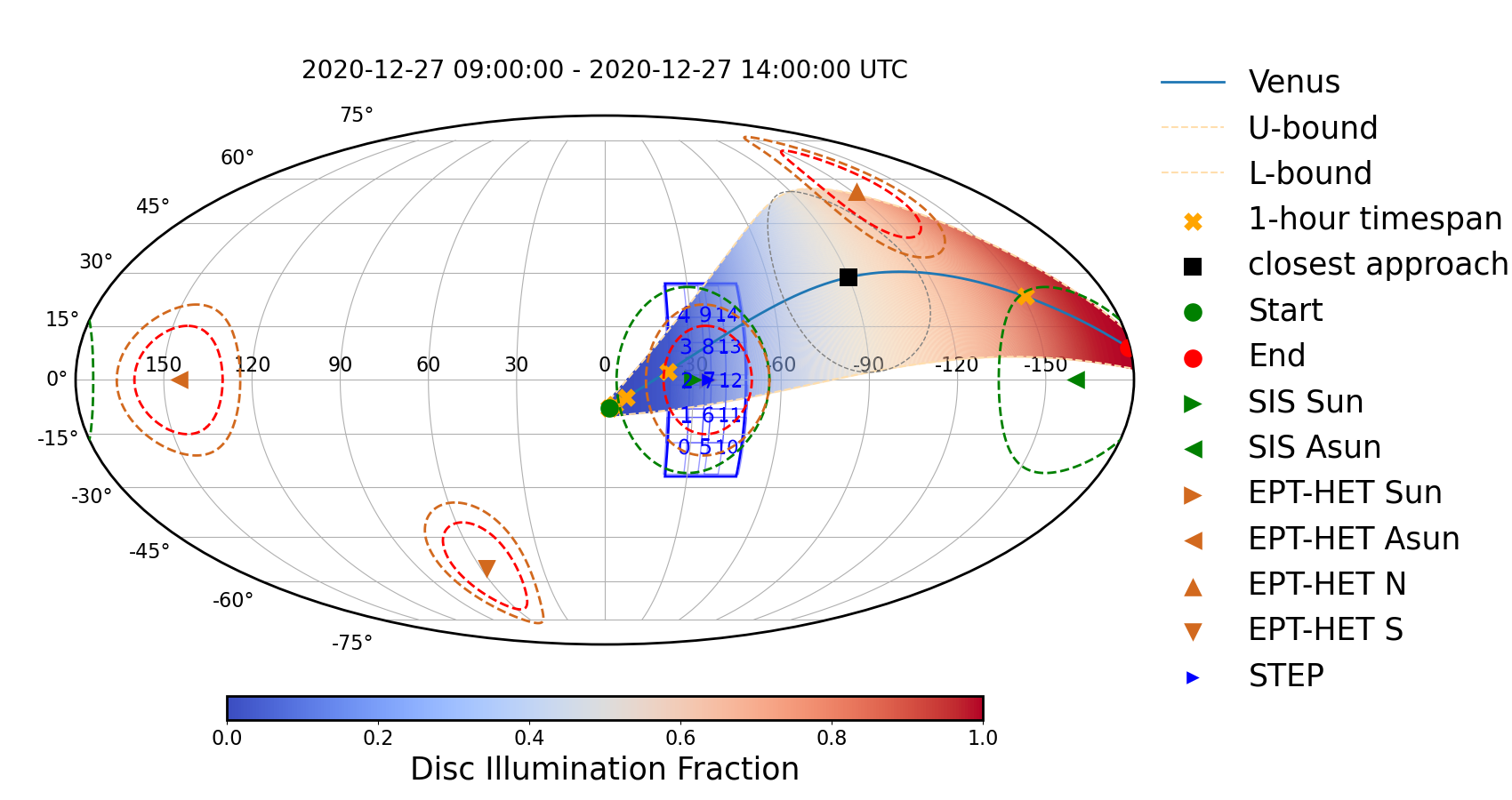}
    \caption{Venus trajectory during the VGAM-1 closest approach as seen from the EPD FOVs in the spacecraft reference frame. The colored band represents the Venus angular size with the fraction of the disc illuminated as seen from the spacecraft. Yellow crosses along the trajectory mark one-hour time intervals. During the maneuver, the spacecraft was rolled 130\degree in order to blind the star-tracker from Venus. }
    \label{fig:vgam_fov}
\end{figure}

\begin{itemize}
    \item[] \textbf{Solar Array (SA) Rotations}: Along the orbit, the SA is tilted in order to sustain the power production and maintain a safe temperature regime. The fixed positions of the SA are safe in terms of EPD science, however, some STEP pixels appear to register stray light while the SA rotates through certain angles \citep[see Fig.~14 in][]{prieto-etal-2021}. The corresponding time periods are flagged in the L2 data files.

\item[]\textbf{Rolls}: Although the spacecraft nominal pointing is Sun-oriented, rotations around the Sun-spacecraft axis are scheduled for several reasons, for example, for calibration purposes, or to change the orientation of the High-gain antenna, or for special maneuvers such as GAMs. Rolls are no risk for EPD but impact the measurements as the sensors change their nominal pointing, in particular, those in the sun-ward direction which point roughly towards the nominal Parker spiral.

\item[]\textbf{Attitude Disturbances}: Trajectory Correction Maneuvers (TCM) and Attitude Maintenance Windows constitute the most important operations for orbit correction. These maneuvers are flight-dynamics related maintenance activities during which thruster firings and departures from the nominal pointing are allowed. During these periods, SIS HV is typically ramped down or reduced.
Depending on the exact maneuver, STEP and EPT/HET can also be switched off to avoid risk to the sensors.
Finally, in this category we also include maneuvers designed to remove ice from the instrument boom. This maneuver is performed with a $210^{\circ}$ pitch slew with the Sun progressively illuminating the spacecraft during which illumination sensitive instruments are made safe as well. This operation is referenced as a de-icing operation in table \ref{app:ops}. 

\item[]\textbf{Gravity Assist Maneuvers (GAM)}: GAM's are a fundamental ingredient in the mission planning, as they are needed in order to get Solar Orbiter close to the Sun and reach the high latitudes of the final orbit. Thruster firings may be needed starting one month ahead of the GAM and may require safety measures for the EPD sensors. The planetary albedo may pose additional risks for thermal loads or stray-light. Moreover, the roll angle may be non-nominal.

Figure~\ref{fig:vgam_fov} shows the Venus trajectory during VGAM-1 as seen from the EPD FOVs. SIS was in safe mode with its high voltage off during a few hours around the encounter and STEP was switched off for 1.5 hours close to the periapsis to avoid stray light from Venus stressing its read-out electronics. Despite this, valuable data were successfully taken with EPD, as described in \cite{Allen2021}
\end{itemize}

A table with a list of the data known issues in the first year of EPD operations as well as the main upcoming planned maneuvers impacting EPD is provided in Appendix \ref{app:ops}.

\section{Summary, conclusions, and outlook}
\label{sec:conclusions}

The first year's worth of data from Solar Orbiter's EPD has shown an exciting richness of results despite a very quiet Sun which is only beginning to exit its last activity minimum. We have illustrated that EPD works as anticipated using the 10-11 December 2020 solar particle event and have also presented a handful of topics which we believe users of EPD data should be aware of. They are summarized in the following brief paragraphs. 

Both STEP and EPT measure electrons and ions, neither of them can discriminate between different elements. That fact is especially important for the low-energy bins of STEP, where heavy solar wind ions can contribute to varying degrees depending on the instantaneous direction of the IMF \citep{nemecek-etal-2020, Wraase18a_SEPT}.

This limitation is, however, also an important effect for EPT. The discrimination between electrons and ions is also not as straightforward as one may presume because of the additional deflection of low-energy ions in the STEP magnet (ion) channel and because ions above several hundred keV can penetrate the polyimide layer on EPT's foil (electron) detector. 

The energy calibration of the lowest energy bins of the HET C BGO calorimeter detector and in the highest energy bins of its B Si-solid-state detectors, which are located just ahead of the BGO crystal, is currently not well understood and is being improved by the EPD team. 

The inter-calibration between SIS and EPT looks very promising but has not yet been finalized. We are eagerly awaiting some large proton-rich solar particle events to allow a more accurate cross calibration. We will only be able to fully characterize the instrument behavior under very high particle flux conditions, when solar activity increases and we observe really large SEP events.

We consider these points to be teething issues that tend to be typical at this point in a mission. The data provided to the scientific community have been corrected for these effects in the best way currently possible, updates will be made as we better understand these instrument peculiarities. Lists of particle intensity enhancements (electrons and ions) as measured by EPT during the first year of Solar Orbiter operations, as well as the different maneuvers or spacecraft operations which may have an effect on the data, are provided in the appendices.

The EPD data are provided at ESA's SOAR\footnote{\url{http://soar.esac.esa.int/soar/}} at the latest 90 days after receipt on Earth. This is not equal to 90 days after the measurement because there is a variable latency in getting nominal data telemetered back to Earth, depending on the location of Solar Orbiter and on the fill state of the spacecraft solid-state memory \citep{Mueller_2020}. Several papers in this special issue have made use of data from SOAR, which demonstrates the community interest. Future changes to instrument settings will be irrelevant to users, as the data provided in the SOAR are in physical units. Yet, these changes will be documented in the updated EPD calibration files in the SOAR.

\begin{acknowledgements}
Solar Orbiter is a space mission of international collaboration between ESA and NASA, operated by ESA. We gratefully acknowledge the hard work and dedication of the many people who contributed to EPD on Solar Orbiter at all of our institutions. It was a unique experience for all of us. We acknowledge the Spanish Ministerio de Ciencia, Innovaci\'on y Universidades for their invaluable support on the
development of the EPD suite and the ICU under grants FEDER/MCIU – Agencia Estatal de Investigaci\'on/Projects ESP2015-68266-R, ESP2017-88436-R and PID2019-104863RB-I00/AEI/10.13039/501100011033. We also thank acknowledge the financial support by the Spanish MINECO-FPI-2016 predoctoral grant with FSE.
We thank the German Federal Ministry for Economic Affairs and Energy and the German Space Agency (Deutsches Zentrum für Luft- und Raumfahrt, e.V., (DLR)) for their unwavering support of STEP, EPT, and HET under grants numbers 50OT0901, 50OT1202, 50OT1702, and 50OT2002; and ESA for supporting the build of SIS under contract number SOL.ASTR.CON.00004, as well as the University of Kiel and the Land Schleswig-Holstein for their support of SIS. We thank NASA headquarters and the GSFC Solar Orbiter project office for support of SIS at APL under contract NNN06AA01C. A. A. acknowledges the support by the Spanish Ministerio de Ciencia e Innovaci\'{o}n (MICINN) under grant PID2019- 105510GB-C31 and through the “Center of Excellence Mar\'{i}a de Maeztu 2020-2023” award to the ICCUB (CEX2019-000918- M). Solar Orbiter magnetometer operations are funded by the UK Space Agency (grant ST/T001062/1). Tim Horbury is supported by STFC grant ST/S000364/1. We acknowledge funding from the European Union’s Horizon 2020 research and innovation programme under grant agreement No. 101004159 (SERPENTINE). We thank E. Valtonen from the University of Turku, Finland, for his inestimable contribution to EPD. We thank the ESA project and operations teams for their expert knowledge and for their dedication and excellent support of the mission, especially during the COVID-19 pandemic. 
\end{acknowledgements}
\bibliographystyle{aa} 
\bibliography{EPDfirstyear}

\begin{appendix} 
\section{Event catalog}
\label{app:events}

\begin{table*}[ht]
\centering
\caption{List of electron intensity enhancements observed by EPT during the first year of observations.}
\label{tab:catalogue_electrons}
\begin{tabular}{cccccccc}
\hline
\hline
\textbf{Start date} &
  \textbf{Onset time} &
  \textbf{Peak date} &
  \textbf{Peak time} &
  \textbf{Peak intensity} &
  \textbf{Max energy} &
  \textbf{dt-Telescope} &
  \textbf{Refs} \\ \midrule
\textbf{(dd/mm/yyyy)} &
  \textbf{(UTC)} &
  \textbf{(dd/mm/yyyy)} &
  \textbf{(UTC)} &
  \textbf{(cm$^2$ s sr MeV)$^{-1}$} &
  \textbf{(keV)} &
  \textbf{(min)} &
   \\ \hline
11/07/2020 & 02:31 & 11/07/2020 & 02:44 & 6.99e2 & 154-167    & 1      & a \\
19/07/2020 & 10:47 & 19/07/2020 & 14:37 & 1.29e2 & 78-85      & 5      & a \\
20/07/2020 & 20:47 & 20/07/2020 & 21:27 & 3,78e2 & 130-142    & 5      & a \\
21/07/2020 & 01:27 & 21/07/2020 & 01:42 & 2.24e2 & 78-85      & 5-Omni & a \\
21/07/2020 & 03:07 & 21/07/2020 & 03:58 & 1.72e3 & 53-58      & 1      & a \\
21/07/2020 & 04:56 & 21/07/2020 & 04:58 & 1.24e3 & 199-218    & 1      & a \\
21/07/2020 & 06:36 & 21/07/2020 & 06:42 & 1.59e3 & 257-281    & 1      & a \\
21/07/2020 & 07:16 & 21/07/2020 & 07:40 & 3.07e3 & 218-237    & 1      & a \\
21/07/2020 & 08:02 & 21/07/2020 & 08:05 & 4.52e3 & 1020-2400\tablefootmark{1,2} & 1      & a \\
22/07/2020 & 23:44 & 22/07/2020 & 23:51 & 1.73e3 & 119-130    & 1      & a \\
22/10/2020 & 12:53 & 22/10/2020 & 14:37 & 1.87e3 & 237-257\tablefootmark{2}     & 1      &   \\
23/10/2020 & 02:32 & 23/10/2020 & 02:37 & 3.49e2 & 78-82      & 5      &   \\
26/10/2020 & 17:52 & 26/10/2020 & 18:22 & 2.03e2 & 167-183    & 5      &   \\
28/10/2020 & 13:17 & 28/10/2020 & 13:27 & 1.38e2 & 85-93      & 5      &   \\
14/11/2020 & 10:15 & 15/11/2020 & 07:15 & 1.29e2 & 62-67      & 30     &   \\
17/11/2020 & 09:43 & 17/11/2020 & 10:26 & 4.26e3 & 1020-2400\tablefootmark{1}   & 1      &   \\
17/11/2020 & 12:45 & 17/11/2020 & 13:05 & 1.27e3 & 85-93      & 2      &   \\
17/11/2020 & 14:45 & 17/11/2020 & 14:53 & 9.43e2 & 85-93      & 2      &   \\
17/11/2020 & 18:37 & 17/11/2020 & 18:45 & 1.96e3 & 73-78      & 1      &   \\
18/11/2020 & 12:09 & 18/11/2020 & 12:17 & 5.75e2 & 73-78      & 1      &   \\
18/11/2020 & 13:29 & 18/11/2020 & 13:39 & 1.31e3 & 85-93      & 1      &   \\
18/11/2020 & 14:35 & 18/11/2020 & 14:42 & 1.57e3 & 85-93      & 1      &   \\
18/11/2020 & 18:39 & 18/11/2020 & 22:07 & 8.02e2 & 93-101     & 1      &   \\
18/11/2020 & 12:09 & 18/11/2020 & 12:17 & 5.75e2 & 58-62      & 1      &   \\
19/11/2020 & 06:13 & 19/11/2020 & 06:47 & 1.12e3 & 78-85      & 2      &   \\
20/11/2020 & 20:15 & 20/11/2020 & 22:01 & 2.37e3 & 1020-2400\tablefootmark{1}   & 1      &   \\
24/11/2020 & 06:51 & 24/11/2020 & 11:00 & 4.78e2 & 93-101     & 2      &   \\
24/11/2020 & 13:21 & 24/11/2020 & 13:52 & 1.10e3 & 101-110    & 1      &   \\
24/11/2020 & 19:32 & 24/11/2020 & 21:25 & 1.17e3 & 434-471\tablefootmark{2}     & 1      &   \\
29/11/2020 & 13:38 & 29/11/2020 & 18:02 & 8.22e3 & 5990-17980\tablefootmark{1}  & 1      & b \\
09/12/2020 & 05:22 & 09/12/2020 & 05:57 & 3.02e2 & 78-85      & 5      &   \\
10/12/2020 & 08:06 & 10/12/2020 & 08:14 & 1.73e3 & 78-85      & 1      &   \\
10/12/2020 & 23:43 & 10/12/2020 & 23:59 & 9.92e4 & 1020-2400\tablefootmark{1}   & 1      &   \\
11/12/2020 & 08:33 & 11/12/2020 & 09:07 & 6.81e2 & 101-110    & 2      &   \\
12/12/2020 & 09:45 & 12/12/2020 & 10:25 & 2.43e2 & 67-73      & 10     &   \\
13/12/2020 & 10:05 & 13/12/2020 & 10:53 & 6.86e2 & 110-119    & 2  &    \\ 
15/02/2021 & 13:37 & 15/02/2021 & 14:03 & 6.750e2 & 101-110   & 2  &     \\
20/02/2021 & 21:45 & 21/02/2021 & 02:35 & 1.613e2 & 119-130   & 10 &    \\ \hline
\end{tabular}
\tablefoot{
   \tablefoottext{1}{Maximum energy range for electrons as observed by HET.}
   \tablefoottext{2}{Possible heavier-species contamination.} \\
   {\bf References.} (a) \citet{gomez-herrero-etal-2021}; (b) \citet{kollhoff-etal-2021}.}
\end{table*}

\begin{table*}[ht]
\centering
\caption{List of ion intensity enhancements observed by EPT during the first year of observations.
}
\label{tab:catalogue_ions}
\begin{tabular}{cccccccc}
\hline
\hline
\textbf{Onset date} &
  \textbf{Onset time} &
  \textbf{Peak date} &
  \textbf{Peak time} &
  \textbf{Peak intensity} &
  \textbf{Max energy} &
  \textbf{dt-Telescope} &
  \textbf{Refs} \\ \midrule
  \textbf{(dd/mm/yyyy)} &
  \textbf{(UTC)} &
  \textbf{(dd/mm/yyyy)} &
  \textbf{(UTC)} &
  \textbf{(cm$^2$ s sr MeV)$^{-1}$} &
  \textbf{(keV)} &
  \textbf{(min)} &
   \\ \hline
18/04/2020 & 02:55 & 18/04/2020 & 04:25 & 3.75e1 & 187-202   & 10      & a \\
20/04/2020 & 09:57 & 20/04/2020 & 10:27 & 4.50e1 & 151-162   & 5       & a \\
07/06/2020 & 19:27 & 08/06/2020 & 01:07 & 8.51e1 & 187-202   & 5       & b \\
19/06/2020 & 03:02 & 19/06/2020 & 04:03 & 1.31e3 & 411-446   & 1       & c,d \\
19/06/2020 & 10:37 & 19/06/2020 & 11:02 & 1.05e2 & 274-298   & 5       & c,d \\
11/07/2020 & 17:07 & 11/07/2020 & 17:17 & 4.19e1 & 218-235   & 5-Omni  & c \\
12/07/2020 & 03:07 & 12/07/2020 & 17:30 & 7.00e1 & 218-235   & 5       &  \\
21/07/2020 & 02:47 & 21/07/2020 & 21:12 & 4.63e2 & 4100-4400 & 5       & c \\
23/10/2020 & 07:32 & 23/10/2020 & 12:57 & 1.03e2 & 804-873   & 5       &  \\
23/10/2020 & 22:12 & 23/10/2020 & 23:52 & 9.75e1 & 804-873   & 5       &  \\
13/11/2020 & 01:09 & 13/11/2020 & 01:12 & 3.75e2 & 202-218   & 1-South &  \\
14/11/2020 & 14:09 & 14/11/2020 & 15:45 & 1.50e2 & 235-254   & 1-South &  \\
15/11/2020 & 23:45 & 15/11/2020 & 13:05 & 7.19e1 & 298-321   & 10-Omni &   \\
18/11/2020 & 03:15 & 18/11/2020 & 09:35 & 9.63e1 & 298-321   & 10      & e \\
19/11/2020 & 00:55 & 19/11/2020 & 03:45 & 1.45e2 & 950-1030  & 10      & e \\
24/11/2020 & 19:03 & 25/11/2020 & 15:59 & 1.37e3 & 4870-5310 & 2-Omni  &   \\
26/11/2020 & 12:21 & 26/11/2020 & 13:57 & 1.19e3 & 1220-1340 & 2       &   \\
29/11/2020 & 14:22 & 29/11/2020 & 15:52 & 7.09e1 & 522-572   & 15      & f \\
30/11/2020 & 02:02 & 30/11/2020 & 04:27 & 1.85e2 & 522-572   & 5       & f \\
30/11/2020 & 09:01 & 30/11/2020 & 10:15 & 7.25e2 & 5830-6130 & 2       & f \\
02/12/2020 & 23:52 & 02/12/2020 & 00:52 & 3.58e1 & 202-218   & 15-Omni &   \\
03/12/2020 & 14:07 & 03/12/2020 & 15:52 & 7.00e1 & 202-218   & 5-North &   \\
09/12/2020 & 10:35 & 09/12/2020 & 03:55 & 9.26e1 & 5830-6130 & 10      &   \\
10/12/2020 & 04:39 & 10/12/2020 & 05:01 & 4.50e2 & 676-739   & 1       &   \\
11/12/2020 & 09:45 & 11/12/2020 & 19:25 & 5.40e2 & 5830-6130 & 10      &   \\
14/12/2020 & 12:27 & 14/12/2020 & 21:01 & 6.13e2 & 411-446   & 2       &   \\
15/12/2020 & 11:22 & 15/12/2020 & 11:22 & 1.33e2 & 141-151   & 5       &   \\
19/12/2020 & 21:32 & 19/12/2020 & 23:47 & 3.79e1 & 162-174   & 5-Omni  &   \\
20/12/2020 & 04:05 & 20/12/2020 & 05:01 & 1.61e2 & 254-274   & 2-Omni  &   \\
02/02/2021 & 15:30 & 02/02/2021 & 16:30 & 1.96e1 & 141-151   & 60      &   \\
04/02/2021 & 14:30 & 04/02/2021 & 17:30 & 2.31e1 & 202-218   & 60      &   \\
15/02/2021 & 21:25 & 16/02/2021 & 00:25 & 7.37e1 & 235-254   & 10      &   \\
21/02/2021 & 22:35 & 22/02/2021 & 03:05 & 1.59e2 & 321-349   & 10      &   \\ \hline
\end{tabular} %
\tablefoot{
   {\bf References.} (a) \cite{vforstner-etal-2021}, \citet{kilpua-etal-2021}; (b) \citet{telloni-etal-2021}; (c) \citet{Mason_2020}; (d) \citet{aran-etal-2021}; (e) \citet{Bucik_21}; (f) \citet{kollhoff-etal-2021}, \citet{Mason-etal-2021b}. }
\end{table*}

We compiled tables of the main energetic particle enhancements measured during the first year of observations, for electrons (Table~\ref{tab:catalogue_electrons}) and ions (Table~\ref{tab:catalogue_ions}). The tables show the onset date and time of the enhancement, the time of the highest intensity (i.e., peak) of the period as well as the value of the peak, an estimated range of the highest energy of the intensity, and the time resolution used for the characterization of each event. In those cases where another telescope than the sunward-pointing one measured the highest intensity, the corresponding telescope is indicated in the dt-telescope column. The last column provides references to studies that investigate the event, if known.

The selection of the events has been performed using a reference energy range constructed by integrating the EPT foil energy bins from 53 keV to 85 keV (seven energy bins in total) for the electrons and the EPT magnet energy bins from 132 keV to 220 keV for the ions (seven energy bins as well). These reference channels were also used to compute the onset time, the peak time and peak intensity of the intensity increases. We included all enhancements rising at least a factor of two above the quiet-time background of these reference channels, as observed by the telescope of EPT, which measures the highest intensity.

The events are selected and checked manually and their onsets are calculated as the moment that the first one-minute channel-averaged intensity exceeds the previous mean intensity by at least two standard deviations and keeps above it for at least two consecutive intervals. Most of them are studied for the Sun telescope. If another telescope is used, this is indicated in the table. In addition, averaging has been applied to some events which were deemed too noisy, the smoothing interval is indicated in column \textit{dt-Telescope}.

The maximum energy range reached is estimated by observing each event in different energy channels without averaging, that is, with the same time resolution for each energy bin. The energy range corresponds to the last energy bin that fulfills the condition of reaching at least a factor of 2 higher than the background intensities. In the case of electrons, when the enhancement was also observed by HET, the highest energy bin of HET that fulfills the criteria is indicated instead. In some cases, the event seems to be perturbed by other species that the sensor is not designed for. Those events are flagged as "possible heavier-species contamination."

We find that over the first year, EPD measured a total of 36 electron events and 29 ion events fulfilling these basic criteria. It is important to note, however, that enhancements at lower energies than those used to prepare these tables will not appear in them because they are not intense enough to fulfill the selection criteria. The corotating interaction region (CIR) events observed by SIS between April-August 2020, and studied by~\citet{allen-etal-2020}, are examples of such enhancements.
Furthermore, we note that this catalog extends over two mission phases: the commissioning phase and cruise phase. For the former, Solar Orbiter data is not fully publicly available as it was intended mainly for commissioning, during which the instrumental settings were changed multiple times.

\section{Operations}
\label{app:ops}

This appendix lists the most important time periods when EPD was affected by spacecraft operations. The list contains the type, start and end times of the maneuvers, and the affected sensor. During the reported periods, some sensors may have been turned off or presented unusual pitch angle coverage or illumination problems, as specified in the comments column. Data provided in the SOAR are flagged accordingly.

\clearpage
\onecolumn

\begin{landscape}
\begin{longtable}{ccccccc} 
\caption{List of spacecraft operations impacting EPD.} \\
\hline
\hline
\textbf{Maneuver} & \textbf{Start Date} & \textbf{Start Time} & \textbf{End Date} & \textbf{End Time} & \textbf{Sensor} & \textbf{Comment} \\ \hline
\endfirsthead
\hline
\multicolumn{7}{c}{\textbf{Continuation of Table}
\ref{tab:ops_table}}\\
\hline
\hline
\textbf{Maneuver} & \textbf{Start Date} & \textbf{Start Time} & \textbf{End Date} & \textbf{End Time} & \textbf{Sensor} & \textbf{Comment} \\ 
\hline
\endhead
\hline
\midrule
\multicolumn{7}{c}{\textbf{continuation on next page}}
\endfoot
\endlastfoot
SA      &       2020-08-30      &       04:26:45        &       2020-08-30      &       04:28:15        &       STEP    &        straylight \\
SA      &       2020-09-30      &       12:31:00        &       2020-09-30      &       12:38:00        &       STEP    &        straylight \\
SA      &       2020-10-11      &       15:02:00        &       2020-10-11      &       15:10:00        &       STEP    &        straylight \\
SA      &       2020-11-03      &       23:48:00        &       2020-11-03      &       23:55:00        &       STEP    &        straylight \\
SA      &       2020-11-11      &       10:00:00        &       2020-11-11      &       10:02:00        &       STEP    &        straylight \\
SA      &       2020-11-11      &       10:27:00        &       2020-11-11      &       10:29:00        &       STEP    &        straylight \\
SA      &       2020-11-24      &       19:29:00        &       2020-11-24      &       19:31:00        &       STEP    &        straylight \\
SA      &       2021-05-02      &       00:05:57        &       2021-05-02      &       00:25:57        &       STEP    &        potential straylight.  \\
SA      &       2021-02-09      &       19:49:50        &       2021-02-09      &       19:59:50        &       STEP    &        potential straylight.  \\
SA      &       2021-04-19      &       08:43:07        &       2021-04-19      &       08:53:07        &       STEP    &        potential straylight.  \\
SA      &       2021-05-25      &       08:56:15        &       2021-05-25      &       09:06:15        &       STEP    &        potential straylight.  \\
SA      &       2021-06-09      &       17:13:40        &       2021-06-09      &       17:23:40        &       STEP    &        potential straylight.  \\
\hline                                                                                          
Attitude Disturbance    &       2020-06-14      &       21:00:18        &       2020-06-15      &       03:00:18        &       SIS     &       SIS HV off\\
Attitude Disturbance    &       2020-07-13      &       08:57:03        &       2020-07-14      &       03:07:03        &       STEP, EPT-HET, SIS    &       SIS HV off. EPT-HET, STEP off. TCM and roll \\
ROLL    &       2020-07-19      &       03:00:00        &       2020-07-19      &       16:16:40        &       EPD     &        12 Full 360 roll for MAG \\
Attitude Disturbance    &       2020-08-10      &       08:00:00        &       2020-08-11      &       02:10:00        &       STEP, EPT-HET, SIS    &       SIS HV off, EPT-HET, STEP off \\
EPD switch off  &       2020-08-13      &       19:00:00        &       2020-08-19      &       11:00:00        &       EPD     &       EPD off due to SSMM issue \\
CSW update      &       2020-09-07      &       00:00:00        &       2020-09-14      &       00:00:00        &       EPD     &        EPD off due to CSW update \\
Attitude Disturbance    &       2020-09-21      &       04:58:48        &       2020-09-21      &       23:08:48        &       STEP, EPT-HET, SIS    &       SIS HV off, EPT-HET, STEP off \\
Attitude Disturbance    &       2020-10-05      &       22:11:31        &       2020-10-06      &       10:11:31        &       STEP, EPT-HET, SIS    &       SIS HV off, EPT-HET, STEP off \\
Attitude Disturbance    &       2020-10-06      &       19:43:35        &       2020-10-07      &       01:43:35        &       SIS     &       SIS HV off \\
Attitude Disturbance    &       2020-11-02      &       06:06:35        &       2020-11-03      &       00:16:35        &       STEP, EPT-HET, SIS    &       SIS HV off, EPT-HET, STEP off. TCM and roll \\
Attitude Disturbance    &       2020-11-04      &       20:35:00        &       2020-11-05      &       03:05:00        &       STEP, EPT-HET, SIS    &       SIS HV off, EPT-HET, STEP off, DEICING \\
EPD Engineering activity        &       2020-11-19      &       10:30:00        &       2020-11-19      &       13:30:00        &       EPD     &       ICU patching\\
VGAM slot       &       2020-11-30      &       00:40:00        &       2020-11-30      &       11:25:00        &       SIS     &       SIS HV off \\
VGAM slot       &       2020-12-01      &       00:00:00        &       2020-12-01      &       03:00:00        &       SIS     &       SIS HV off \\
VGAM slot       &       2020-12-07      &       07:45:00        &       2020-12-07      &       10:45:00        &       SIS     &       SIS HV off \\
VGAM slot       &       2020-12-13      &       00:05:00        &       2020-12-13      &       03:05:00        &       SIS     &       SIS HV off \\
VGAM slot       &       2020-12-13      &       17:26:00        &       2020-12-14      &       06:35:00        &       SIS     &       SIS HV off \\
VGAM slot       &       2020-12-20      &       00:05:00        &       2020-12-20      &       03:05:00        &       SIS     &       SIS HV off \\
VGAM slot       &       2020-12-20      &       17:32:00        &       2020-12-21      &       10:33:00        &       SIS     &       SIS HV off \\
VGAM slot       &       2020-12-23      &       00:05:00        &       2020-12-23      &       03:05:00        &       SIS     &       SIS HV off \\
VGAM slot       &       2020-12-24      &       06:40:00        &       2020-12-24      &       20:30:00        &       SIS     &       SIS HV off \\
VGAM slot       &       2020-12-26      &       23:40:00        &       2020-12-27      &       18:39:00        &       SIS     &       SIS HV off \\
ROLL            &       2020-12-27      &       08:00:00        &       2021-01-03      &       23:00:00        &       EPD     &       $130^{\circ}$ roll for VGAM \\
VGAM closest approach   &       2020-12-27      &       11:30:00        &       2020-12-27      &       13:00:00        &         STEP    &        STEP off \\
VGAM slot       &       2020-12-30      &       00:05:00        &       2020-12-30      &       03:05:00        &       SIS     &       SIS HV off \\
\hline                                                                                          
Attitude Disturbance    &       2021-01-01      &       00:05:00        &       2021-01-01      &       03:05:00        &       SIS     &        SIS HV off \\
Attitude Disturbance    &       2021-01-01      &       18:33:00        &       2021-01-02      &       14:30:00        &       SIS     &        SIS HV off \\
ROLL    &       2021-01-07      &       16:00:00        &       2021-01-08      &       05:16:40        &       EPD     &       12 Full 360 roll for MAG \\
Attitude Disturbance    &       2021-01-11      &       07:52:57        &       2021-01-12      &       03:40:00        &       SIS     &       SIS HV off \\
CSW update      &       2021-01-17      &       20:52:37        &       2021-01-24      &       21:10:43        &       EPD     &       Payload off due to CSW update\\
ROLL    &       2021-02-03      &       09:42:31        &       2021-02-03      &       10:32:31        &       EPD     &        1 Roll from 0 to -20 \\
ROLL    &       2021-02-03      &       19:57:14        &       2021-02-03      &       20:47:14        &       EPD     &        1 Roll from -20 to 0  \\
ROLL    &       2021-02-05      &       09:51:35        &       2021-02-05      &       10:41:35        &       EPD     &        1 Roll from 0 to 45 \\
ROLL    &       2021-02-05      &       20:06:15        &       2021-02-05      &       20:56:15        &       EPD     &        1 Roll from 45 to 0  \\
Attitude Disturbance    &       2021-02-08      &       04:00:00        &       2021-02-08      &       23:07:42        &       SIS     &       SIS HV off \\
ROLL    &       2021-02-18      &       00:00:00        &       2021-02-18      &       16:34:22        &       EPD     &       2 Rolls [0,180] [180,0] for RS calibration\\
ROLL    &       2021-02-23      &       00:00:00        &       2021-02-23      &       07:17:28        &       EPD     &        8 rolls [0,335] in 45 steps for RS calibration\\
Attitude Disturbance    &       2021-03-08      &       08:30:28        &       2021-03-09      &       02:50:00        &       SIS     &       SIS HV off \\
Attitude Disturbance    &       2021-04-05      &       04:00:00        &       2021-04-05      &       23:41:37        &       SIS     &       SIS HV off \\
Attitude Disturbance    &       2021-05-03      &       04:00:00        &       2021-05-03      &       10:00:00        &       SIS     &       SIS HV off \\
Attitude Disturbance    &       2021-05-03      &       15:41:43        &       2021-05-04      &       03:41:43        &       SIS     &       SIS HV off \\
Attitude Disturbance    &       2021-06-01      &       01:34:32        &       2021-06-01      &       11:34:32        &       STEP, EPT-HET, SIS    &       SIS HV off, EPT-HET, STEP off \\
Attitude Disturbance    &       2021-06-03      &       03:06:59        &       2021-06-03      &       15:06:59        &       STEP, EPT-HET, SIS    &       SIS HV off, EPT-HET, STEP off  \\
\hline                                                                                          
EPD switch off  &       2020-03-21      &       12:00:00        &       2020-04-21      &       12:00:00        &       EPD     &       Covid measures. Payload off \\
EPD Engineering activity        &       2020-04-29      &       02:00:00        &       2020-04-29      &       03:30:00        &       STEP, EPT-HET &       Upload new config tables STEP, EPTHET \\
EDP Engineering activity        &       2020-05-12      &       01:00:00        &       2020-05-12      &       02:30:00        &       STEP, EPT-HET &       Upload new config tables EPT-HET \\
EPD switch off  &       2020-05-21      &       12:00:00        &       2020-05-29      &       09:00:00        &       EPD     &       Payload off (SpW overload) \\
EMC test        &       2020-06-02      &       06:00:00        &       2020-06-02      &       23:00:00        &       STEP, EPT-HET &        EPT-HET and STEP on-off  \\
EMC test        &       2020-06-03      &       06:00:00        &       2020-06-03      &       13:00:00        &       STEP, EPT-HET &        EPT-HET and STEP on-off  \\
\hline
\hline
\label{tab:ops_table}
\end{longtable}
\end{landscape}

\end{appendix}
\end{document}